\documentclass[12pt, draftclsnofoot, onecolumn]{IEEEtran}

\newtheorem{theo}{Theorem}

\newtheorem{cor}{Corollary}

\IEEEoverridecommandlockouts

\makeatletter
\newcommand{\biggg}[1]{{\hbox{$\left#1\vbox to 20.5pt{}\right.\n@space$}}}
\newcommand{\Biggg}[1]{{\hbox{$\left#1\vbox to 23.5pt{}\right.\n@space$}}}
\newcommand{\bigggg}[1]{{\hbox{$\left#1\vbox to 26.5pt{}\right.\n@space$}}}
\newcommand{\Bigggg}[1]{{\hbox{$\left#1\vbox to 29.5pt{}\right.\n@space$}}}
\newcommand{\biggggg}[1]{{\hbox{$\left#1\vbox to 32.5pt{}\right.\n@space$}}}
\newcommand{\Biggggg}[1]{{\hbox{$\left#1\vbox to 35.5pt{}\right.\n@space$}}}
\newcommand{\bigggggg}[1]{{\hbox{$\left#1\vbox to 38.5pt{}\right.\n@space$}}}
\newcommand{\Bigggggg}[1]{{\hbox{$\left#1\vbox to 41.5pt{}\right.\n@space$}}}
\makeatother

\usepackage{bm,cite,algorithm,algorithmic,float,amsmath,amssymb}
\usepackage[dvips]{graphicx}
\usepackage{subfigure,amsfonts}
\usepackage{mdwmath}
\usepackage{mdwtab}
\usepackage{xcolor,color}
\usepackage{cool}
\usepackage{balance}
\usepackage{diagbox}
\floatname{algorithm}{Algorithm}

\makeatletter
\renewcommand\paragraph{\@startsection{paragraph}{4}{\z@}%
            {-2.5ex\@plus -1ex \@minus -.25ex}%
            {1.25ex \@plus .25ex}%
            {\normalfont\normalsize\itshape}}
\makeatother
\usepackage{epstopdf}

\usepackage[style=0]{mdframed}
\usepackage{showexpl}

\mdfdefinestyle{exampledefault}{%
rightline=true,innerleftmargin=10,innerrightmargin=10,
frametitlerule=true,frametitlerulecolor=green,
frametitlebackgroundcolor=yellow,
frametitlerulewidth=2pt}

\allowdisplaybreaks

\begin{document}


\title{Cognitive Radio-Inspired Rate-Splitting Multiple Access for Semi-Grant-Free Transmissions}

\author{
Hongwu~Liu, Kyeong Jin Kim, Theodoros A. Tsiftsis,
Bruno Clerckx, \\
Kyung Sup Kwak,
and H. Vincent Poor

\thanks{H. Liu is with the School of Information Science and Electrical Engineering, Shandong Jiaotong University, Jinan 250357, China (e-mail: liuhongwu@sdjtu.edu.cn).}
\thanks{K. J. Kim is with Mitsubishi Electric Research Laboratories, Cambridge, MA 02139 USA (e-mail: kkim@merl.com).}
\thanks{T. A. Tsiftsis is with the School of Intelligent Systems Science and Engineering, Jinan University, Zhuhai 519070, China (e-mail: theo\_tsiftsis@jnu.edu.cn).}
\thanks{B. Clerckx is  with the Department of Electrical and Electronic Engineering, Imperial College London, London, UK (e-mail: b.clerckx@imperial.ac.uk).}
\thanks{K. S. Kwak is with Inha University, Incheon, South Korea (email: kskwak@inha.ac.kr).}
\thanks{H. V. Poor is with the Department of Electrical and Computer Engineering, Princeton University, Princeton, NJ 08544 USA (e-mail: poor@princeton.edu).}
}

\maketitle
\setcounter{page}{1}
\begin{abstract}
In this paper, we propose a cognitive radio-inspired rate-splitting multiple access (CR-RSMA) scheme  to assist semi-grant-free (SGF) transmissions in which a grant-based user (GBU) and multiple grant-free users (GFUs) access the base-station (BS) by sharing the same resource block.
Using the cognitive radio principle, the GBU and admitted GFU are treated as the primary and secondary users, respectively, and rate-splitting is applied at the admitted GFU to realize SGF transmissions.
The admitted GFU's transmit power allocation, target rate allocation, and successive interference cancellation decoding order at the BS are jointly optimized to attain the maximum achievable rate for the admitted GFU without deteriorating the GBU's outage performance compared to orthogonal multiple access.
Due to the extended non-outage zone,  CR-RSMA-assised SGF (CR-RSMA-SGF) transmissions achieve a lower outage probability than SGF transmissions assisted by cognitive radio-inspired non-orthogonal multiple access.
Exact expressions and asymptotic analysis for the admitted GFU's outage probability are derived to evaluate the system performance achieved by  CR-RSMA-SGF transmissions. The superior outage performance and full multiuser diversity gain achieved by   CR-RSMA-SGF transmissions are verified by the analytical and simulation results.
\end{abstract}

\begin{IEEEkeywords}
Rate-splitting, multiple access, grant-free transmissions, outage probability.
\end{IEEEkeywords}

\section{Introduction}

The proliferation of Internet-of-Things (IoT) applications, intelligent robots, and Industry 4.0 networks is resulting in unprecedented need for massive and spectrally efficient connections.
As part of the sixth-generation (6G) evolution roadmap, extremely reliable and low-latency
communication (ERLLC), further-enhanced mobile broadband
(FeMBB), and ultra-massive machine-type communication
(umMTC) have been proposed to enable envisioned  heterogeneous 6G applications \cite{6G_Wireless_Networks}.
For ubiquitous scenarios of umMTC and ERLLC, where traditional grant-based (GB) transmissions are impractical due to the corresponding excessive signaling overhead and massive computational resource consumption, grant-free (GF) transmissions have been proposed, attracting considerable interest from academia and industry \cite{Next_IoT, Sparse_GF_IoT, SCMA_blind}.
A main advantage of GF transmissions is that terminals can   access the network without engaging in lengthy handshaking, whereas the amount of handshaking signaling can exceed the amount of data sent by terminals in GB transmissions. Although GF transmissions are a promising solution for massive
connectivity, GB transmissions cannot be overlooked due to the stringent quality of service (QoS) requirements
of existing grant-based GB users (GBUs).  For this reason, semi-grant-free (SGF) transmissions have been proposed to accommodate the coexistence of GB and GF transmissions, which results in higher spectral efficiency than admitting only GBUs or GF users (GFUs) \cite{SGF_NOMA_Simple}. To opportunistically admit GFUs to resource blocks occupied by GBUs, non-orthogonal multiple access (NOMA)-assisted SGF (NOMA-SGF) transmissions have been proposed, which eliminates complex handshaking processes for admitting GFUs \cite{SGF_NOMA_Simple,SGF_NOMA_QoS}. By using NOMA, user collisions caused by the GFUs' contention can be resolved by spectrum sharing with the aid of successive interference cancellation (SIC).

To avoid system performance degradation of a GBU in NOMA-SGF transmissions, the GFU's contention must be managed appropriately taking into account the GBU's QoS requirements. Specifically, outage performance experienced by the GBU should be comparable to that of its orthogonal multiplex access (OMA) counterpart \cite{SGF_NOMA_Simple,SGF_NOMA_QoS}. In  \cite{SGF_NOMA_Simple}, a distributed contention protocol was proposed to guarantee that a fixed number of GFUs are admitted. It was seen that this open-loop protocol suffered from user collisions at a rate similar to that seen in pure GF transmissions. In \cite{SGF_NOMA_QoS},
the received GBU's signal power at the base station (BS) was used to determine an interference threshold, which was then broadcast to GFUs to facilitate distributed contentions.
Considering that existing power-domain NOMA schemes rely on superposition coding and transmit power allocation, advanced SIC can be used to maintain a high transmission reliability while preventing the GBU from unnecessary awareness of the GFUs' contention \cite{SGF_NOMA_QoS,SGF_NOMA_Geometry_Conferece,
SGF_NOMA_Advanced}.
In \cite{Unveiling_SIC_PartI}, hybrid SIC was proposed to decode the desired signals based on their relative power levels, i.e., the admitted GFU's signal can be decoded at the first or second stage of SIC to attain the allowed maximum achievable rate, which can be used in conjunction with transmit power allocation to further improve the reliability of NOMA-SGF transmissions \cite{SGF_NOMA_PA}.

When viewed in the context of underlay cognitive radio (CR), the paired NOMA users can be regarded as a primary user (PU) and a secondary user (SU), respectively \cite{NOMA_pairing}, which signifies that GB and GF transmissions in NOMA-SGF systems can be treated as primary and secondary transmissions, respectively.
Since CR-inspired NOMA (CR-NOMA) with SIC can achieve only a  subset on the capacity region boundary of uplink multiple access channels (MACs),  the outage performance of SU in this setting can result in decreased user fairness \cite{RS_CP_SC}.
To avoid deterioration in the PU's outage performance, both PU and SU can be allowed to access the same  BS simultaneously in CR-NOMA,  and a new hybrid SIC  was proposed to improve the achievable rate of the SU in such a system \cite{Hybrid_SIC_NOMA_New}.
To achieve arbitrary points on the capacity region boundary of uplink MACs, rate-splitting multiple access (RSMA) proposed in \cite{Rate_Splitting_MAC} can also be implemented  in a CR-inspired way, which can extend the non-outage zone for CR-NOMA \cite{RS_CP_SC}.
In \cite{New_RS_CR_NOMA}, the authors proposed a CR-inspired RSMA (CR-RSMA) scheme to improve the SU's outage performance by admitting both PU and SU simultaneously to the same BS.
However, the admission contention from multiple GFUs was not taken into account in  CR-RSMA  and the effects of the GFUs' contention on CR-RSMA-assisted SGF (CR-RSMA-SGF) transmissions are still not known.

\subsection{Related Work}

{\emph{1)  RSMA:} RSMA has received significant attention recently due to its capability to improve the spectral/energy efficiencies, robustness, reliability, and latency of downlink and uplink multi-user transmissions \cite{RSMA_Survey_Trends, RSMA_6G_PartI}.
In downlink RSMA, the users' messages are split into common and private streams using  available channel state information at the transmitter (CSIT) \cite{RS_LTE}. By treating inter-user interference  flexibly, i.e., interference can be partially decoded and partially treated as noise, downlink RSMA not only bridges space division multiple access (SDMA) and NOMA, but also achieves superior system performance, so that it is regarded as a promising enabling technology for 6G new radio (NR) \cite{RS_EURASIP,Sum_rate_RS_approach,RS_unifying, RSMA_new_frontier} (see also \cite{RSMA_Survey_Trends} and references therein).

However,  RSMA and the corresponding applications for uplink MACs are still in their infancy. Being essentially different from downlink RSMA, which decodes  common and private streams partially at a receiver to balance the decoding performance and complexity, the split data streams from all the users are fully decoded at the BS using SIC in uplink RSMA \cite{RS_SUM_RATE, Rate_splitting_Coding_NOMA, Rate_splitting_NOMA, RS_NOMA_maxmin_fairness}.
In \cite{RS_SUM_RATE}, the sum-rate maximization problem was investigated for uplink RSMA in which the proportional rate constraints among users were considered. To enhance the outage performance of uplink NOMA, several rate-splitting  schemes were proposed in \cite{Rate_splitting_Coding_NOMA, Rate_splitting_NOMA, RS_NOMA_maxmin_fairness}.
For single-input multiple-output (SIMO) NOMA, uplink rate-splitting was investigated to  guarantee max-min user fairness in \cite{RS_NOMA_maxmin_fairness}.
Nevertheless, an exhaustive search was needed to find the optimal SIC decoding order and optimal power allocation for uplink RSMA \cite{RS_SUM_RATE, RS_NOMA_maxmin_fairness}.
Using the CR principle,  adaptive power allocation and rate-splitting were proposed to improve the outage performance and user fairness for uplink NOMA in \cite{RS_CP_SC}.

Recently, uplink RSMA has been  applied to realize physical layer network slicing for ultra-reliable and low-latency (URLLC) and enhanced mobile broadband communications (eMMB) \cite{RSMA_URLLC_eMBB}.
Also,  RSMA has been used to
support  URLLC and  eMMB in 6G NR downlink transmissions \cite{RSMA_URLLC_downlink}.
In \cite{Performance_RSMA}, the outage performance of uplink RSMA was investigated taking into account all possible SIC decoding orders. In \cite{RS_NOMA_User_Cooperation},  cooperative RSMA was proposed to realize uplink user cooperation. In \cite{RS_Aerial_Networks} and \cite{RS_Uplink_NOMA_Satellite}, rate-splitting schemes were designed for uplink aerial networks and satellite communications, respectively.
Rate-splitting applications for cell-free machine-type communications and device-to-device fog radio access networks were investigated in \cite{RS_CellFree_MTC} and \cite{RS_D2D_FRAC}, respectively.

{\emph{2)  NOMA-SGF transmissions:} To control the number of admitted GFUs for NOMA-SGF transmissions, two contention mechanisms, namely open-loop contention and distributed contention protocols, have been  proposed in \cite{SGF_NOMA_Simple}, in which an interference temperature alike channel gain threshold was broadcast to aid the GFUs' contention. Furthermore, a dynamic interference threshold proportional to the received GBU's signal power was proposed in \cite{SGF_NOMA_Geometry_Conferece} to determine the admitted GFUs, which results in reduced GFUs' interference compared to the open-loop contention.
The authors in \cite{SGF_NOMA_QoS} and \cite{Unveiling_SIC_PartI}  proposed CR-NOMA with hybrid SIC decoding order to enhance  transmission reliability of the admitted GFU, meanwhile ensuring that the GBU experiences the same outage performance as in OMA. An ergodic rate analysis was provided for CR-NOMA-assisted SGF (CR-NOMA-SGF) transmissions in \cite{SGF_NOMA_Ergodic_Rate}.
To  efficiently leverage the capability of CR-NOMA to achieve  the capacity region of uplink MACs, adaptive power allocation with hybrid and fixed SIC decoding orders were proposed  in \cite{SGF_NOMA_Advanced} and \cite{SGF_NOMA_PA}.
Further, adaptive power allocation was proposed in \cite{SGF_NOMA_Performance_UL} for CR-NOMA-SGF transmissions to improve the outage performance and sum rate.
In \cite{mMIMO_SGF}, a downlink collided-preamble feedback was applied to facilitate massive multiple-input multiple-output assisted SGF random access. For tactile IoT networks that use NOMA-SGF transmissions, joint power allocation and sub-channel assignment was investigated in \cite{SGF_NOMA_Tactile}. In \cite{Power_levels_SGF_NOMA}, a user barring
scheme was proposed to increase the average arrival rate for NOMA-SGF transmissions with multiple transmit power levels.
In \cite{SGF_NOMA_MA_DRL} and \cite{Power_Pool_GF_NOMA},  multi-agent deep reinforcement learning (MADRL) was applied to optimize the transmit power for  NOMA-SGF and NOMA-assisted GF (NOMA-GF) transmissions. Moreover, MADRL was used to optimize transmit power allocation, sub-channel assignment and reflection beamforming for an intelligent reflecting surface (IRS) aided NOMA-SGF system \cite{SGF_NOMA_MADRL_IRS}.
\vspace{-0.10in}

\subsection{Motivation and Contributions}

Although NOMA-SGF transmissions can accommodate the coexistence of  GB and GF transmissions, this system achieves only a subset of the uplink MAC capacity region as noted above. In contrast,  the full capacity region of uplink MACs can be achieved by using RSMA
\cite{Rate_Splitting_MAC}. However, to maximize the sum-rate and ensure user fairness, exhaustive search was used to determine the optimal SIC decoding order and optimal power allocation for uplink RSMA, which is computationally prohibitive
\cite{RS_SUM_RATE, RS_NOMA_maxmin_fairness}. For delay-limited transmissions, the authors in \cite{RS_NOMA_UL_fair} proposed a CR-inspired transmit power allocation to aid rate-splitting, in which target rates of the SU's data streams were chosen heuristically.

On the other hand, the recently proposed CR-RSMA
can improve the SU's outage performance \cite{New_RS_CR_NOMA} and reduce the task-offloading latency for mobile edge computing \cite{RSMA_MEC} by optimizing the transmit power and target rate allocations.
Motivated by the capacity region achieved by SGF transmissions, we propose a new CR-RSMA that adopts RSMA and CR principles to assist SGF transmissions, with the aim of improving the system performance under the uplink RSMA framework. In CR-RSMA-SGF transmissions, rate-splitting is conducted at the admitted GFU, while the optimal transmit power and target rate allocations at the admitted GFU and SIC decoding order at the BS are jointly optimized using the CR principle, which improves the admitted GFU's outage performance significantly.

The main contributions of this paper are summarized as follows:
\begin{itemize}
\item A new CR-RSMA is proposed to allow simultaneous access of the GBU and GFU to the BS without deteriorating the outage performance of the GBU compared to OMA. Applying  CR-inspired rate-splitting at the admitted GFU, the SIC decoding order, transmit power allocation, and target rate allocation are jointly optimized for the admitted GFU to obtain the maximum achievable rate. By exploiting the full capacity region achieved by RSMA, the non-outage zone of CR-RSMA-SGF transmissions is extended significantly compared to CR-NOMA-SGF transmissions.
\item With respect to the capacity region, the extended non-outage zone and reduced outage zone are described for CR-RSMA-SGF transmissions in comparison with the CR-NOMA-SGF transmissions.
    Due to the extended non-outage zone, additional target rate pairs containing the higher target rates for the admitted GFU can be supported by CR-RSMA-SGF transmissions, so that the outage performance of the admitted GFU is significantly improved when CR-RSMA is applied.
\item We derive an exact expression for the outage probability of the admitted GFU and its approximation in the high signal-to-noise ratio (SNR) region. These analytical results reveal that the full multiuser diversity gain is achieved by CR-RSMA-SGF transmissions. Moreover, benefiting from the obtained maximum achievable rate for the admitted GFU, the multiuser diversity can be achieved in a wider target rate regions than those of the existing CR-NOMA-SGF transmissions.
\item Various computer simulation results are presented to verify the accuracy of the derived analytical results and high SNR approximations.
    The superior outage performance of the admitted GFU achieved by CR-RSMA-SGF transmissions is verified by these simulation results. The impact of the target rate and number of GFUs on the outage performance of the admitted GFU are revealed.
\end{itemize}

The remainder of this paper is organized as follows: Section II presents the system model and CR-RSMA-SGF transmissions, respectively; In Section III, the outage performance of the admitted GFU achieved by CR-RSMA-SGF transmissions is analyzed and the high SNR approximation for the outage probability is derived; In Section IV, simulation results are presented for corroborating the superior outage performance of CR-RSMA-SGF transmissions, and Section V summarizes this work.

\section{System Model and CR-RSMA Transmissions}



\subsection{System Model}

We assume that the SGF transmission system consists of multiple GBUs and multiple GFUs. To prevent GF transmissions from generating too much interference to  GBUs, all the GFUs and GBUs are divided into multiple groups, each of which contains $K$ GFUs and one GBU. In each group, we assume that only one out of $K$ GFUs is paired with the GBU for simultaneous uplink transmissions, whereas multiple GFUs from different groups can be simultaneously admitted through wireless resource allocation among the groups.

Without loss of generality, we consider SGF transmissions within a single group, in which the GBU and the $k$th GFU are denoted by $U_0$ and $U_{\tilde k} \in \{U_{\tilde 1}, U_{\tilde 2},  \cdots, U_{\tilde K} \}$, respectively. The channel coefficients from $U_0$ and $U_{\tilde k}$ to the BS are denoted by $h_0$ and $h_{\tilde k}$ (${\tilde k}={\tilde 1}, {\tilde 2}, \cdots, {\tilde K}$), respectively, which are modeled as independent and identically distributed (i.i.d.) circular symmetric complex Gaussian random variables with zero mean and unit variance. We assume that the channels follow a quasi-static fading, i.e., the channel coefficients remain constant during a single transmission block and can vary from one transmission block to another independently. Moreover, the   channel gains can be ordered as
\begin{eqnarray}
|h_1|^2 \le |h_2|^2 \le \cdots \le |h_K|^2,     \label{eq:h_order}
\end{eqnarray}
where the subscripts denote the ordered indices.
It should be noted that the above ordering information is unavailable to  the $K$ GFUs and BS.
In the considered system, we assume that the $K$ GFUs have the knowledge of their own channel state information (CSI) and the admitted GFU's CSI is not required to be known at the BS prior to the SGF transmissions. In addition, the BS has acquired the information of the GBU's CSI  and transmit power.

Without loss of generality, we assume that the $\ell$th GFU $U_\ell$ is admitted to transmit ($1 \le \ell \le K$).
For each block of transmissions, $U_\ell$ can split its message signal $\bar x_\ell$ into two parts $\bar x_{{\ell,1}}$ and $\bar x_{{\ell,2}}$. Corresponding to the message signals  $\bar x_\ell$, $\bar x_{{\ell,1}}$, and $\bar x_{{\ell,2}}$, the generated transmit signals at $U_\ell$ are denoted by $x_\ell$, $x_{{\ell,1}}$, and $x_{{\ell,2}}$, respectively.
In each transmission block, $U_0$ and $U_\ell$ simultaneously transmit their signals to the BS. Then, the received signal at the BS can be written as:
\begin{eqnarray}
y = \sqrt{P_0} h_0 x_0 + \sqrt{\alpha P_s} h_\ell x_{{\ell,1}} + \sqrt{(1-\alpha)P_s} h_\ell x_{{\ell,2}} + w, \label{eq:y}
\end{eqnarray}
where $P_0$ and $P_s$ denote  the transmit power of $U_0$ and $U_\ell$, respectively, $x_0$ is the transmit signal of $U_0$, $\alpha$ is the transmit power allocation factor at $U_\ell$ satisfying $0 \le \alpha \le 1$, and $w$ is additive white Gaussian noise (AWGN) at the BS with zero mean and unit variance.
We assume that each transmit signal $\tilde x \in \{x_0, x_\ell,  x_{{\ell,1}}, x_{{\ell,2}} \}$ is coded by an independent Gaussian code book and satisfies ${\mathbb{E}}\{|\tilde x|^2\} = 1$, where ${\mathbb{E}}\{\cdot\} $ is the expectation operator.

At the BS, the decoding order $x_{{\ell,1}} \to x_0 \to x_{{\ell,2}}$ is adopted in SIC to recover the signal $x_{\ell,1}$, $x_0$,  and $x_{\ell,2}$, sequentially. According to the uplink RSMA principle \cite{Rate_Splitting_MAC}, the SIC decoding order $x_{{\ell,1}} \to x_0 \to x_{{\ell,2}}$ ensures that the full capacity region boundary of MACs can  be achieved. Then, the received signal-to-interference-plus-noise ratio (SINR) (or signal-to-noise ratio (SNR)) for decoding $x_{{\ell,1}}$, $x_0$,  and $x_{{\ell,2}}$  can be expressed as follows:
\begin{eqnarray}
    \gamma_{\ell,1} =   \frac{\alpha P_s |h_\ell|^2}{ P_0|h_0|^2 + (1-\alpha)P_s |h_\ell|^2 + 1} ,   \label{eq:SINR_rs1}
\end{eqnarray}
\begin{eqnarray}
    \gamma_{0} =   \frac{P_0|h_0|^2}{(1-\alpha)P_s |h_\ell|^2 + 1} , \label{eq:SINR_GB}
\end{eqnarray}
and
\begin{eqnarray}
    \gamma_{\ell,2} =   (1-\alpha)P_s |h_\ell|^2. \label{eq:SINR_rs1}
\end{eqnarray}
For the admitted $U_0$ and $U_\ell$, the achievable rate are given by $R_{0} = \log_2(1 + \gamma_{0})$ and $R_{{\ell}} =  R_{\ell,1} + R_{\ell,2}$, respectively, with $ R_{\ell,1} = \log_2(1 + \gamma_{\ell,1})$ and $ R_{\ell,2} = \log_2(1 + \gamma_{\ell,2})$ representing the achievable rates to   transmit $x_{\ell,1}$ and $x_{\ell,2}$, respectively.

{\textbf{\emph{Remark 1}}}: By simply interchanging $x_{{\ell,1}}$ and $x_{{\ell,2}}$, all possible SIC decoding orders at the BS can be categorized into three types as $x_{{\ell,1}}  \to x_{{\ell,2}} \to x_0$, $x_0 \to x_{{\ell,1}} \to x_{{\ell,2}}$, and $x_{{\ell,1}} \to x_0 \to x_{{\ell,2}}$. Using the logarithmic product law, we can readily prove that applying  $x_{{\ell,1}}  \to x_{{\ell,2}} \to x_0$ is equal to applying $x_{\ell} \to x_0$ in the sense of attaining the achievable rates $R_\ell$ and $R_0$. In such a case, rate-splitting is unnecessary, so does the decoding order $x_0 \to x_{{\ell,1}} \to x_{{\ell,2}}$. As it will be seen in the next subsection,  NOMA with the SIC decoding orders $x_{\ell} \to x_0$ and $ x_0 \to x_{\ell}$ cannot approach the full capacity region boundary of MACs. Thus, the decoding order $x_{{\ell,1}} \to x_0 \to x_{{\ell,2}}$ is used in the CR-RSMA-SGF transmissions with respect to its capability to achieve the full capacity region boundary.

\subsection{CR-RSMA-SGF Transmissions}

For the considered CR-RSMA-SGF transmissions, to guarantee that the GBU $U_0$ achieves the same outage performance as in OMA, a CR analogous interference threshold is broadcast to the $K$ GFUs to aid user contention.
Let $\hat R_0$, $\hat R_s$, $\hat R_{s,1}$, and $\hat R_{s,2}$ denote the target rates to transmit $x_0$, $x_\ell$, $x_{\ell,1}$, and $x_{\ell,2}$, respectively. In addition, we define $\hat R_s = \hat R_{s,1} + \hat R_{s,2}$, $\hat R_{s,1}  = \beta \hat R_s$, and $\hat R_{s,2}  = (1-\beta) \hat R_s$, where $0 \le \beta \le 1$ is the target rate allocation factor.
With respect to the SIC decoding order $x_{{\ell,1}} \to x_0 \to x_{{\ell,2}}$, the signal $x_0$ can be decoded correctly only when the following constraints are satisfied, i.e.,
\begin{eqnarray}
    R_{\ell,1} \ge \hat R_{s,1}  {\text{~~and~~}}   R_{0} \ge \hat R_{0}. \label{eq:constraint1}
\end{eqnarray}
Under the constraints in \eqref{eq:constraint1},
the GBU  achieves the same outage
performance as in OMA
when $\Pr\left(R_{\ell,1} \ge \hat R_{s,1} \right) = 1$ and
\begin{eqnarray}
\Pr \left(R_0 \ge \hat R_{0} \right) = \Pr\left(\log_2(1+ P_0|h_0|^2) \ge \hat R_{0} \right). \label{eq:constraint3}
\end{eqnarray}
By substituting \eqref{eq:SINR_GB} into  \eqref{eq:constraint3}, the equality in \eqref{eq:constraint3} holds under the condition of
\begin{eqnarray}
(1-\alpha)P_s |h_\ell|^2 \le  \hat\tau , \label{eq:constraint4}
\end{eqnarray}
where $\hat \tau  = \tfrac{P_0|h_0|^2}{2^{\hat R_0}-1} - 1$. The equation \eqref{eq:constraint4} indicates that
the interference power caused by the transmission of $\sqrt{(1-\alpha)P_s} h_\ell x_{{\ell,2}}$ cannot surpass $\hat \tau$ to ensure the correct decoding of $x_{0}$ at the second stage of SIC.
By assuming that $\Pr\left(R_{\ell,1} \ge \hat R_{s,1} \right) = 1$ is guaranteed in the  CR-RSMA-SGF transmissions, as it will be explained later in this subsection, the interference threshold to be broadcast by the BS is  determined as \cite{SGF_NOMA_QoS}:
\begin{eqnarray}
\tau = \max\left\{0, \hat\tau \right\}. \label{eq:tau}
\end{eqnarray}

The admission procedure of the CR-RSMA-SGF transmissions is presented as follows:

\begin{itemize}
    \item The BS broadcasts the pilot signals to assist the users to estimate CSI.
    \item $U_0$ feeds back its CSI and $P_0$ to the BS.
    \item The BS calculates $\tau$ according to \eqref{eq:tau} and broadcasts it to the $K$ GFUs.
	\item Each GFU calculates the achievable rate using its own CSI and the corresponding optimal $\alpha^*$ and $\beta^*$, which will be  provided in the later part of this subsection.
	\item Through the distributed contention, the GFU that obtains the maximum achievable rate is admitted by the BS.
\end{itemize}

With respect to the capability of uplink RSMA to achieve the full capacity region boundary of MACs, the design goal of the CR-RSMA-SGF transmissions is to maximize the achievable rate for the admitted GFU meanwhile guaranteeing that the GBU achieves the same outage performance as in OMA. With respect to all the possible CSI realizations, the rate-splitting operation and associated $\alpha^*$ and $\beta^*$ of the CR-RSMA-SGF transmissions are jointly designed as follows:

{\bf{\emph{Case I:}}} \emph{$0< P_s |h_K|^2 \le \tau$.} In this case, the maximum interference level caused by admitting an arbitrary GFU is not greater than the interference threshold $\tau$. When $U_\ell$ is admitted, the maximum achievable rate
is given by $R_{\ell} = R_{\ell,2} =  \log_2(1+\gamma_{\ell,2})$ considering that $x_{\ell,2}$ is interference-freely decoded at the last stage of  SIC. In other words, $U_\ell$ will allocate all of $P_s$ to transmit $x_\ell$ by setting $x_{\ell,2} = x_\ell$. Consequently, $U_K$, which has the greatest channel gain among all the GFUs, is admitted to attain the maximum achievable rate $R_K = R_{K,2}  = \log_2(1+\gamma_{K,2})$.
In this case, the optimal transmit power and target rate allocation factors are respectively given by
\begin{eqnarray}
\alpha^* = 0  {\text{~~and~~}}  \beta^* = 0,
\end{eqnarray}
where $(\cdot)^*$ denotes the optimal solution for the corresponding parameter.

Since only $x_{K,2}$ ($x_{K,2}=x_K $) is transmitted, the SIC decoding order $x_{K,1} \to x_0 \to x_{K,2}$ degenerates to $x_0 \to x_{K}$. Consequently, the achievable rate  of the admitted  GFU can be expressed as
\begin{eqnarray}
R_K^{(\rm I)} =  \log_2\left(1 + P_s |h_K|^2  \right).
\end{eqnarray}

{\bf{\emph{Case II:}}} \emph{$0 < \tau < P_s |h_1|^2$ or $P_s |h_{k}|^2 < \tau < P_s |h_{k+1}|^2$ with $k=1, 2, \ldots, K-1$.} In this case, it can be seen that $\hat \tau >0 $. Then, a GFU $U_{\ell}$   whose channel gain is larger than $\hat \tau$ will be admitted, where $k+1 \le \ell \le K$.
Due to $\hat \tau > 0$, the GBU's signal $x_0$ can be correctly decoded at the second stage of SIC ($x_{\ell,1} \to x_0 \to x_{\ell,2}$) subject to the constraints $R_{\ell,1} > \hat R_{\ell,1}$ and
$R_{\ell, 2} \le  \log_2(1+ \hat \tau)$. Thus, the admitted GFU's achievable rate $R_{\ell} = R_{\ell,1} + R_{\ell,2}$ can be maximized by first maximizing $R_{\ell, 2} =  \log_2(1+\gamma_{\ell,2})$, which is obtained as $R_{\ell, 2} =  \log_2(1+ \hat \tau)$ by setting $\gamma_{\ell, 2} = \hat \tau$. The  corresponding optimal transmit power allocation factor is given by
\begin{eqnarray}
\alpha^*  = 1 -  \frac{\hat \tau}{P_s |h_{\ell}|^2}.
\end{eqnarray}
By setting the target rate $\hat R_{\ell, 2} = \log_2(1+ \hat\tau)$ to transmit $x_{\ell,2}$, the optimal target rate allocation factor is given by
\begin{eqnarray}
\beta^*   = 1 -  \frac{\log_2(1 + \hat \tau)}{\hat R_s}.
\end{eqnarray}
Furthermore, $R_{\ell,1}$ can be maximized by setting $\ell = K$ considering that $U_K$ has the greatest channel gain, i.e., the achievable rate for the GFU is maximized by admitting $U_K$.
Accordingly, the optimal transmit power and target rate allocation factors are respectively given by $\alpha^*(K)$ and $\beta^*$, and the achievable rate in this case can be written as
\begin{eqnarray}
R_K^{({\rm II})}  &\!\!\! =\!\!\! & R_{K,1}^{({\rm II})} + R_{K,2}^{({\rm II})},
\end{eqnarray}
where
\begin{eqnarray}
R_{K,1}^{({\rm II})}  &\!\!\! =\!\!\! &  \log_2\left( 1 + \frac{P_s |h_K|^2 - \hat\tau}{P_0 |h_0|^2 + \hat \tau + 1} \right)
\end{eqnarray}
and
\begin{eqnarray}
R_{K,2}^{({\rm II})}  &\!\!\! =\!\!\! & \log_2(1 + \hat \tau).
\end{eqnarray}

Since the received SINR/SNR  $\gamma_0(\alpha^*(K))$ and $\gamma_{K,2}(\alpha^*(K))$ provide the necessary conditions for the correct decoding of  $x_0$ and $x_{K,2}$ in SIC processing $x_{{K,1}} \to x_0 \to x_{{K,2}}$, only the failure decoding of $x_{K,1}$ can result in error propagation in SIC. To avoid this error propagation, $R_{K,1}^{({\rm II})} \ge \hat R_{s,1}$ is required to decode $x_{K,1}$ correctly or equivalently $R_K^{({\rm II})} \ge \hat R_{s}$; Otherwise, both $U_0$ and $U_K$ encounter outage. Therefore, in the proposed CR-RSMA-SGF transmissions, $U_K$ is permitted to transmit only when $R_{K,1}^{({\rm II})} \ge \hat R_{s,1}$; Otherwise, $U_K$ keeps silence while $U_0$ is transmitting alone to the BS.

{\bf{\emph{Case III:}}} \emph{$\tau = 0$.} In this case, it can be seen that $\hat \tau < 0$, so that $x_0$ cannot be  correctly decoded due to a weak channel gain $|h_0|^2$. To avoid error propagation in SIC processing $x_{\ell,1} \to x_0 \to x_{\ell,2}$ caused by the failure decoding of $x_0$, the admitted GFU is chosen not to transmit $x_{\ell,2}$ rather to transmit $x_\ell$ by setting $x_{\ell,1} = x_\ell$ and using the whole $P_s$.
Therefore, the optimal transmit power and target rate allocation factors are given by
\begin{eqnarray}
\alpha^* =1  {\text{~~and~~}}  \beta^* = 1.
\end{eqnarray}
Since $U_K$ has the greatest channel gain among the GFUs, it is admitted to attain the maximum achievable rate as
\begin{eqnarray}
R_K^{(\rm III)} =  \log_2\left(1 + \frac{P_s |h_K|^2}{P_0 |h_0|^2 + 1} \right)
\end{eqnarray}
and the corresponding SIC decoding order $x_{K,1} \to x_0 \to x_{K,2}$ degenerates to $x_K \to x_0$.

{\textbf{\emph{Remark 1}}}: Due to the channel gain ordering $|h_1|^2 \le |h_2|^2 \le \cdots \le |h_K|^2$, the maximum achievable rates in Cases I, II, and III for the admitted GFU are always attained by admitting $U_K$, no matter which decoding order is applied in SIC. Since the channel coefficients of all the $K$ GFUs follow i.i.d. complex Gaussian distribution, all the GFUs have the equal probability to have the greatest  channel gain, or equivalently, to be admitted. Thus, the admission probability for all the $K$ GFUs are equal in the proposed CR-RSMA-SGF transmissions.

{\textbf{\emph{Remark 2}}}:
In Cases I and III, the CR-NOMA-SGF transmissions also admit $U_K$ and apply the SIC decoding order $x_0 \to x_K$ and $x_K \to x_0$, respectively, to obtain the maximum achievable rates \cite{SGF_NOMA_QoS}. Nevertheless, in Case II,  the CR-NOMA-SGF transmissions admit $U_K$ (or $U_k$) subject to $P_s |h_{1}|^2 > \tau$ (or $P_s |h_{k}|^2 < \tau < P_s |h_{k+1}|^2$). For the CR-NOMA-SGF transmissions, the achievable rate of the admitted GFU in Case II is given by \cite{SGF_NOMA_QoS}
\begin{eqnarray}
R_{\rm NOMA}^{(\rm II)} =  \left\{ {\begin{array}{*{20}{c}}
{R_K^{(\rm III)},}&{0 < \tau < P_s |h_1|^2} \\
{\max\big\{R_k^{(\rm I)}, R_K^{(\rm III)}\big\},}&{P_s |h_{k}|^2 < \tau < P_s |h_{k+1}|^2}
\end{array}} \right. ,
\end{eqnarray}
where $k=1, 2, \ldots, K-1$.
Since $ R_K^{(\rm II)} > R_K^{(\rm III)} $ and $ R_K^{(\rm II)} > R_k^{(\rm I)} $ ($k=1, 2, \ldots, K-1$) hold for $\tau > 0$, we have $R_K^{({\rm II})} >  R_{\rm NOMA}^{(\rm II)}$, i.e., the CR-RSMA-SGF transmissions  always achieve  a larger achievable rate for the admitted GFU than that of the CR-NOMA-SGF transmissions in Case II.

{\textbf{\emph{Remark 3}}}: In Case II, $U_K$ only transmits its signal when $R_{K,1}^{({\rm II})} \ge \hat R_{s,1}$, which prevents error propagation in SIC processing $x_{K,1} \to x_0 \to x_{K,2}$. When $R_{K,1}^{({\rm II})} < \hat R_{s,1}$, $U_K$ keeps silence and only the GBU $U_0$ transmits. Thus, the GBU $U_0$ does not encounter outage when the CR-RSMA-SGF transmissions operate in Case II. As such, the outage events occur to the transmissions of $x_0$ in Cases I and III are the same as in OMA, and the GBU $U_0$ achieves the same outage performance as in OMA when the CR-RSMA-SGF transmissions are applied.

\section{Outage Performance Analysis}

In the CR-RSMA-SGF transmissions, since the GBU's outage performance is guaranteed to be the same as in OMA, we mainly focus on the outage performance of the admitted GFU. In this section, we first introduce the non-outage zone to clarify the advantage of the CR-RSMA-SGF transmissions. Then, we derive the analytical expression for the outage probability in closed-form  and investigate the asymptotic outage performance in the high SNR region.

For the uplink MACs in which the two-user $U_0$ and $U_K$ are admitted simultaneously, the non-outage zone is defined by
\begin{eqnarray}
\bar{\cal{O}}  \triangleq \left\{  \{\hat R_0, \hat R_s  \} \big| ~ \hat R_0 \le R_0, \hat R_s \le R_K  \right\}. \label{eq:non_outage_zone}
\end{eqnarray}
If a target rate pair $\{\hat R_0, \hat R_s  \}$ lies in the non-outage zone, both $U_0$ and $U_K$ can attain the achievable rates $R_0 \ge \hat R_0$ and $R_K \ge \hat R_K$ such that neither $U_0$ nor $U_K$ encounters the outage event.
Otherwise, both $U_0$ and $U_K$ are in outage. Corresponding to \eqref{eq:non_outage_zone}, the set $ {\cal{O}}  \triangleq \left\{  \{\hat R_0, \hat R_s  \} \big| ~ \hat R_0 > R_0, \hat R_s > R_K  \right\}$ is called the outage zone when both $U_0$ and $U_K$ are admitted simultaneously.

\begin{figure}[tb]
    \begin{center}
    \includegraphics[width=4.4in]{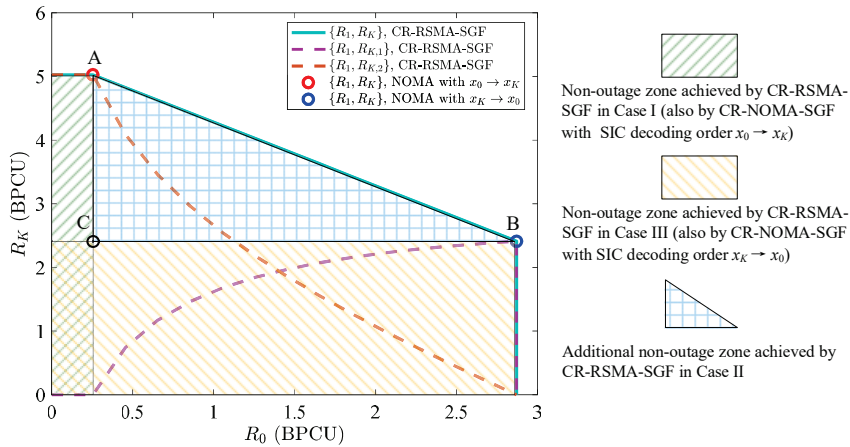}
    \caption{Non-outage zones achieved by different schemes ($P_0|h_0|^2 = 8$ dB, $P_s|h_K|^2 = 15$ dB).}
    \label{fig:subfig3c}
    \end{center}
    \vspace{-0.3in}
\end{figure}

In Fig. 1, an example of the non-outage zone achieved by the CR-RSMA-SGF transmissions is illustrated. For the comparison purpose,  the non-outage zones achieved by the CR-NOMA-SGF transmissions are also illustrated in which the SIC decoding orders $x_0 \to x_K$ and $x_K \to x_0$ are respectively utilized \cite{SGF_NOMA_QoS}. In this example, we assume $0 < \tau < P_s |h_1|^2$.
From Fig. 1, we can see that the non-outage zones achieved by the CR-NOMA-SGF transmissions using the SIC decoding orders $x_0 \to x_K$ and $x_K \to x_0$ are identical to those achieved by the  CR-RSMA-SGF transmissions in Cases I and III, respectively.
Thus, in Cases I and III, the CR-RSMA-SGF transmissions can achieve the same outage performance as that of the CR-NOMA-SGF transmissions. Nevertheless, the CR-NOMA-SGF transmissions cannot prevent $U_0$ and $U_K$ from being in outage when the target rate pair $\{\hat R_0, \hat R_K \}$ lies in the triangle ABC, which is beyond the non-outage zone of the CR-NOMA-SGF transmissions.
Fortunately, the points on the line AB in Fig. 1 can be achieved by rate-splitting. Specifically, an additional non-outage zone, the triangle ABC, is achieved when the CR-RSMA-SGF transmissions  operate  in Case II.
Thus, the non-outage zone of the CR-RSMA-SGF transmissions is extended compared to that of the CR-NOMA-SGF transmissions, which shows that the CR-RSMA-SGF transmissions can support more target rate pairs than that of the CR-NOMA-SGF transmissions.

With respect to the operations of the CR-RSMA-SGF transmissions in Cases I, II, and III, the outage probability of the admitted GFU can be written as
\begin{eqnarray}
P_{\rm out} = P_{\rm out}^{(\rm I)} + P_{\rm out}^{(\rm II)} + P_{\rm out}^{(\rm III)},
\end{eqnarray}
where $P_{\rm out}^{(\rm I)}$, $P_{\rm out}^{(\rm II)}$, and $P_{\rm out}^{(\rm III)}$ respectively denote the probability of that $U_K$ encounters outage in Cases I, II, and III, which can be expressed as
\begin{eqnarray}
    P_{\rm out}^{(\rm I)} = \Pr\left( 0< P_s |h_K|^2 <  \tau, R_K^{(\rm I)} < \hat R_s  \right), \label{eq:Pout_I}
\end{eqnarray}
\begin{eqnarray}
    P_{\rm out}^{(\rm II)} =  \sum\limits_{k=0}^{K-1} {P_{\rm out}^{(\rm II,\it{k})}} ,
\end{eqnarray}
and
\begin{eqnarray}
    P_{\rm out}^{(\rm III)} = \Pr\left( \tau = 0, R_K^{(\rm III)} < \hat R_s  \right) \label{eq:Pout_III}
\end{eqnarray}
 with
\begin{eqnarray}
    P_{\rm out}^{(\rm II,0)} = \Pr\left( 0 < \tau < P_s |h_1|^2, R_K^{(\rm II)} < \hat R_s  \right)  \label{eq:Pout_II_0}
\end{eqnarray}
and for $k=1, 2, \ldots, K-1$,
\begin{eqnarray}
    P_{\rm out}^{(\rm II,\it{k})} = \Pr\left( P_s |h_{k}|^2 < \tau < P_s |h_{k+1}|^2, R_K^{(\rm II)} < \hat R_s  \right).
\end{eqnarray}

The following theorem provides an exact expression for the admitted GFU's outage probability  achieved by the CR-RSMA-SGF transmissions.

\begin{theo}
    Assume that $K \ge 2$, the outage probability of the admitted GFU is given by
    \begin{eqnarray}
        P_{\rm out}  &\!\!\!=\!\!\!& \frac{\varphi_0}{K(K-1)}  \sum\limits_{n=0}^{K}  \binom{K}{n}   (-1)^n  ~ \mu_{1} \nu(0, \mu_{2})  \nonumber \\
        &\!\!\! \!\!\!& +  \sum\limits_{k=1}^{K-2}  \varphi_k  \sum\limits_{m=0}^{K-k}   \binom{K\!-\!k}{m} (-1)^m \sum\limits_{n=0}^{k} \binom{k}{n} (-1)^{n}e^{\frac{n}{P_s}}   \mu_3 \nu(n,  \mu_4) \nonumber \\
        &\!\!\! \!\!\!& + \frac{ \varphi_0}{K-1} \sum\limits_{n=0}^{K-1} \binom{K\!-\!1}{n}  (-1)^n e^{\frac{n}{P_s} }     \left( e^{\frac{1}{P_s}} \nu(n, \mu_5)  - e^{-\frac{\epsilon_0 + \epsilon_s + \epsilon_0\epsilon_s  }{P_s} } \nu(n, \mu_6)   \right)  \nonumber \\
        &\!\!\! \!\!\!& + \sum\limits_{n = 0}^K \binom{K}{n} (-1)^n e^{\frac{n}{P_s}} \nu(n, 0)   + \left(1 - e^{-\eta_s}  \right)^K e^{- \eta_0 (1+\epsilon_s)  }    \nonumber \\
        &\!\!\! \!\!\!& + \sum\limits_{n = 0}^K \binom{K}{n}  (-1)^n e^{-n \eta_s} \frac{1 - e^{-(1+n \eta_s P_0 )\eta_0 }}{1 + n \eta_s  P_0},   \label{eq:Pout_GFU}
    \end{eqnarray}
    where $\epsilon_0 \triangleq 2^{\hat R_0}-1$, $\epsilon_s \triangleq 2^{\hat R_s} - 1  $, $\eta_0 \triangleq \frac{\epsilon_0}{P_0}$,  $\eta_s \triangleq \frac{\epsilon_s}{P_s}$, $\mu_{1} = e^{\frac{K - n(1+\epsilon_0)(1+\epsilon_s)}{P_s}}$, $\mu_{2} = \frac{K-n}{P_s \eta_0} - \frac{n P_0}{P_s}$, $ \mu_3 = e^{\frac{K-k - m \left(1+\epsilon_0 \right) \left(1+ \epsilon_s \right) }{P_s}}$, $ \mu_4 = \frac{K-k-m}{P_s \eta_0 }  - \frac{m P_0}{P_s}  $, $\mu_5 = \frac{1}{P_s\eta_0}$, $\mu_6 = -\frac{P_0}{P_s}$, $\varphi_0 = \frac{K!}{(K-2)!}$, $\varphi_k = \frac{K!}{k! (K-k)!}$ for $1 \le k \le K-2$, and
    \begin{eqnarray}
    \nu(n, \mu) = \left\{ {\begin{array}{*{20}{c}}
        {\epsilon_s  \eta_0  }, &{ {\rm if~~} \mu = -1 - \frac{n}{P_s\eta_0} },\\
        {\frac{e^{- \eta_0 \left(\frac{n  }{P_s\eta_0}  + \mu + 1\right) } - e^{- \eta_0 (1+\epsilon_s)  \left(\frac{n  }{P_s\eta_0}  + \mu + 1\right)  }}{  \frac{n }{P_s\eta_0}  + \mu + 1  } }, &{\text{otherwise}}.
        \end{array}} \right.
    \end{eqnarray}
\end{theo}
\begin{IEEEproof}
    See Appendix A.
\end{IEEEproof}

\begin{figure}[tb]
    \begin{center}
    \includegraphics[width=4.2in]{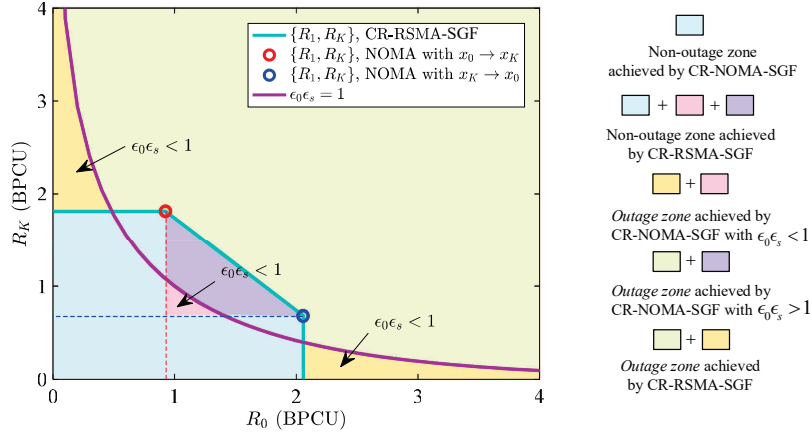}
    \caption{Outage and non-outage zones achieved by different schemes ($P_0|h_0|^2 = 5$ dB, $P_s|h_K|^2 = 4$ dB).}
    \label{fig:subfig3c}
    \end{center}
    \vspace{-0.3in}
\end{figure}

{\textbf{\emph{Remark 4:}}} In deriving the expressions for $P_{\rm out}^{(\rm I)} $ and $P_{\rm out}^{(\rm II)} $, as derived in Appendix A, all the outage events reflect the fact that the upper bound on the channel gain $|h_K|^2$ should be greater than the lower bound on a specific GFU's channel gain, which is always and naturally guaranteed by $|h_0|^2<\eta_0(1+\epsilon_s)$ without imposing additional constraints on $\epsilon_0$ and $\epsilon_s$, so that the derived expression for the outage probability is applicable to all the feasible $\epsilon_0$ and $\epsilon_s$.  On the contrary, for the CR-NOMA-SGF transmissions, the analytical expression for the outage probability in \cite{SGF_NOMA_QoS} is applicable only for $\epsilon_0 \epsilon_s < 1$. As an example, the outage and non-outage zones achieved by the CR-RSMA-SGF and CR-NOMA-SGF transmissions are illustrated in Fig. 2. It can be seen that the outage zone constrained by $\epsilon_0 \epsilon_s < 1$ is a small portion of the whole outage zone for the CR-NOMA-SGF transmissions and a similar phenomenon happens to the CR-RSMA-SGF transmissions as well. Therefore, the analytical results provided in Theorem 1 are more general due to its applicability to all the feasible values of $\epsilon_0$ and $\epsilon_s$. In contrast to the CR-RSMA-SGF transmissions, the CR-NOMA-SGF transmissions result in a worse outage performance due to the extended outage zone.

\begin{theo}
    Assuming that $K \ge 2$, the outage probability experienced by the admitted GFU can be approximated in the high SNR region as follows:
    \begin{eqnarray}
        P_{\rm out}   &\!\!\!\approx\!\!\!& \frac{\varphi_0 \epsilon_0 (1+\epsilon_0)^K}{P_s^{K+1}K(K-1)}   \sum\limits_{n = 0}^K \binom{K}{n} \frac{(-1)^n}{n+1}    \left( (1+\epsilon_s)^{K + 1}  -  (1+\epsilon_s)^{K-n}  \right) \nonumber \\
        &\!\!\! \!\!\!& +  \frac{\varphi_k \epsilon_0  (1+\epsilon_0)^{K-k} (-1)^k }{ P_s^{K+1} } \sum\limits_{m=0}^{K-k} \binom{K-k}{m} (-1)^m (1 + \epsilon_s )^{K-k-m} \nonumber \\
    &\!\!\! \!\!\!&    \times  \sum\limits_{n=0}^{k} \binom{k}{n} (-1)^n
    \frac{(1 + \epsilon_s )^{m+n+1} - 1}{m+n+1} +
    \frac{\varphi_0 \epsilon_0 \epsilon_s^K (1 + \epsilon_0)(1+ \epsilon_s)  }{ P_s^{K+1}K(K-1)}
    \nonumber \\
    &\!\!\! \!\!\!&  -
    \frac{ \varphi_0  \epsilon_s^K (\epsilon_0^{-1} + 1) (K(1+\epsilon_s) + 1) }{P_s^{K+1}K(K-1)(K+1)}
    +   \frac{\epsilon_0 \epsilon_s^{K+1} }{(K+1)P_s^{K+1} }  + \frac{\epsilon_s^K}{P_s^K}  - \frac{\epsilon_0 \epsilon_s^K(1+\epsilon_s)}{P_s^{K+1}}
        \nonumber \\
        &\!\!\! \!\!\!&   +   \frac{\epsilon_s^K \left( (1 + \epsilon_0)^{K+1} -1 \right) }{P_s^{K+1} (K+1)}  - \frac{\epsilon_s^K \left( (\epsilon_0(K+1) -1 )(1+\epsilon_0)^{K+1} +1  \right) }{P_s^{K+2} (K+2) (K+1)}. \label{eq:Pout_app}
    \end{eqnarray}
\end{theo}
\begin{IEEEproof}
    See Appendix B.
\end{IEEEproof}

From the results in Theorem 2, we can see that there is one term in \eqref{eq:Pout_app} being  proportional to $\frac{1}{P_s^K}$, while the other terms are proportional to $\frac{1}{P_s^{K+1}}$ or $\frac{1}{P_s^{K+2}}$. Therefore, we have the following corollary.

\begin{cor}
    Assuming that $K \ge 2$ , the admitted GFU's  outage probability  can be further approximated as $\frac{\epsilon_s^K}{P_s^K}$ in the high SNR region. Thus, a diversity gain of $K$ is achieved by the CR-RSMA-SGF transmissions.
\end{cor}

When the considered system consists of a single GFU and GBU, the two users are paired directly.  Then, the GFU's outage probability achieved by the CR-RSMA-SGF transmissions can be written as
\begin{eqnarray}
    P_{\rm out}  &\!\!\!=\!\!\!& \Pr \left( 0< P_s |h_1|^2 \le \tau,  R_{1}^{(\rm I)}< \hat R_s \right)
    + \Pr \left( 0< \tau< P_s|h_1|^2, R_{1}^{(\rm II)}< \hat R_s \right)  \nonumber \\
    &\!\!\!  \!\!\!&  + \Pr \left(\tau = 0,   R_{1}^{(\rm III)}< \hat R_s \right).
         \label{eq:Pout_single}
\end{eqnarray}

\begin{cor}
    Assume that $K = 1$, the admitted GFU's outage probability is given by
    \begin{eqnarray}
        P_{\rm out}  &\!\!\!=\!\!\!& 1 - e^{-\frac{\epsilon_0 + \epsilon_s+ \epsilon_0 \epsilon_s}{P_s}} \nu(0, \mu_6)
        - e^{-\eta_s - \eta_0(1+\epsilon_s) }
        - \frac{e^{-\eta_s} (1 - e^{-\eta_0 -\epsilon_0\eta_s }) }{1+P_0\eta_s}
    \end{eqnarray}
    and its approximation in the high SNR region is given by
    \begin{eqnarray}
        P_{\rm out}  &\!\!\!\approx \!\!\!&  \epsilon_sP_s^{-1}.
    \end{eqnarray}
\end{cor}

\begin{IEEEproof}
    A proof can be found in \cite{New_RS_CR_NOMA}.
\end{IEEEproof}

{\textbf{\emph{Remark 7:}}} Corollaries 1 and 2 demonstrate that the CR-RSMA-SGF transmissions ensure an achievable diversity gain proportional to the number of the GFUs without resulting in an outage floor.

\section{Simulation Results}

In this section, we present simulation results to verify the accuracy of the analytical results and the superior outage performance of the CR-RSMA-SGF transmissions. For the purpose of comparisons, the admitted GFU's outage probabilities achieved by the CR-NOMA-SGF transmissions \cite{SGF_NOMA_QoS} and
the CR-NOMA-SGF transmissions with power control (CR-NOMA-SGF-PC) \cite{Hybrid_SIC_NOMA_New} are also presented.
In the simulation, all the channel coefficients $h_{\tilde k}$s are randomly generated according to i.i.d. circular symmetric complex Gaussian random variables with zero mean and unit variance and the transmission data rate is measured in bits per channel use (BPCU).

\begin{figure}[tb]
    \begin{center}
    \includegraphics[width=3.17in]{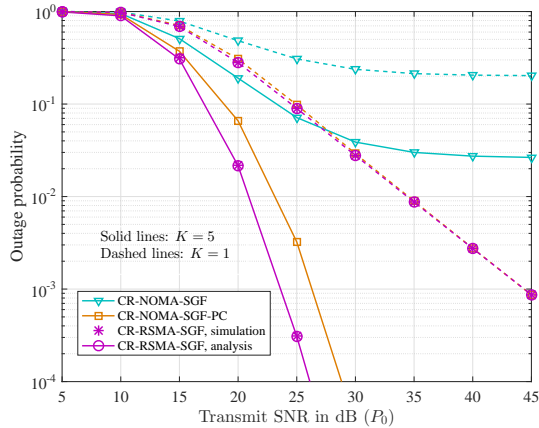}
    \caption{Outage probability comparison of the SGF schemes with $\hat R_0 = 2.5$ BPCU, $\hat R_s = 1.5$ BPCU, and $P_s = \frac{P_0}{15}$.}
    \label{fig:subfig1}
    \end{center}
    \vspace{-0.35in}
\end{figure}

\begin{figure}[tb]
    \begin{center}
    \includegraphics[width=3.17in]{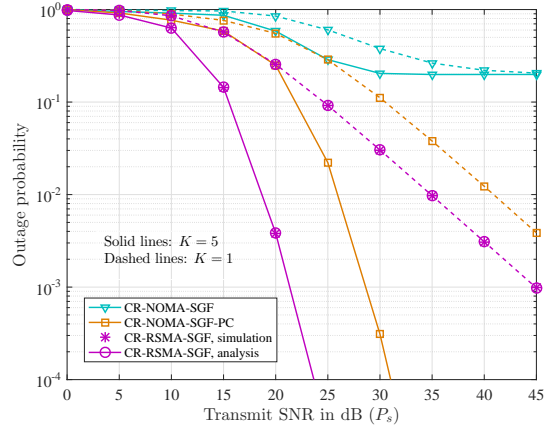}
    \caption{Outage probability comparison of the SGF schemes with $\hat R_0 = 3$ BPCU, $\hat R_s = 3$ BPCU, and the fixed $P_0 = 15$ dB.}
    \label{fig:subfig1}
    \end{center}
    \vspace{-0.35in}
\end{figure}

The outage performance achieved by the  CR-RSMA-SGF transmissions  is compared with those of the CR-NOMA-SGF and CR-NOMA-SGF-PC schemes in Fig. 3, where we set $\hat R_0 = 2.5$ BPCU, $\hat R_s = 1.5$ BPCU, and $P_s = \frac{P_0}{15}$, i.e., the transmit SNR of $U_0$ is 11.76 dB higher than that of the GFUs.
From Fig. 3, we can see that the CR-RSMA-SGF transmissions achieve the smallest  outage probabilities among the three schemes for both $K = 1$ and $K = 5$. The results in Fig. 3 also verify the accuracy of the derived analytical expressions. As transmit SNR increases, the outage probabilities achieved by both the CR-RSMA-SGF and CR-NOMA-SGF-PC schemes decrease monotonically, whereas an outage probability floor occurs for the CR-NOMA-SGF transmissions in the high SNR regime. Although CR-NOMA-SGF-PC scheme avoids the outage probability floor, the achieved outage probability is still higher than that of the CR-RSMA-SGF transmissions. Therefore, the results in Fig. 3 verify that the CR-RSMA-SGF transmissions achieves the superior outage performance compared to the CR-NOMA-SGF and CR-NOMA-SGF-PC schemes.

\begin{figure}[tb]
    \begin{center}
    \includegraphics[width=3.17in]{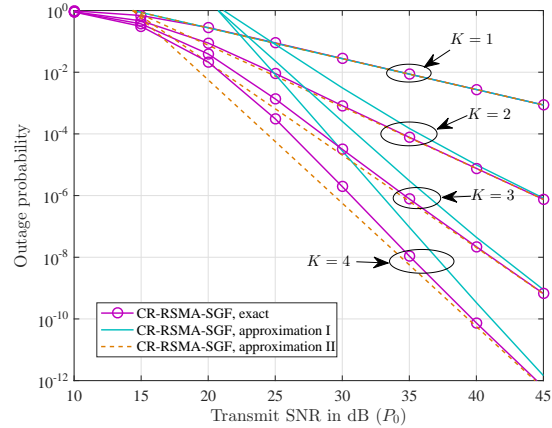}
    \caption{Accuracy of the derived analytical expressions.}
    \label{fig:subfig1}
    \end{center}
    \vspace{-0.35in}
\end{figure}

In Fig. 4, we investigate the outage performance achieved by the SGF schemes for various transmit powers. In particular,
we set a fixed transmit power $P_0 = 15$ dB and vary $P_s$ from 0 dB to 45 dB, which reflects that the GFUs can have greater and lower transmit SNRs than that of $U_0$. The results in Fig. 4 also verify the accuracy of the derived analytical expressions. From Fig. 4, we can see that the CR-RSMA-SGF transmissions achieve the smallest outage probabilities in the whole SNR region. As transmit SNR increases, the outage probability achieved by the CR-RSMA-SGF transmissions decreases monotonically for both $K=1$ and $K=5$. Also, the CR-NOMA-SGF transmissions achieve the worst outage probabilities in the whole SNR region. Especially in the high SNR region, the outage probability floor occurs for the CR-NOMA-SGF transmissions. The results in Fig. 4 verify that the superior outage performance of the CR-RSMA-SGF transmissions is irrespective of whether the GFUs are stronger users or weaker users compared to $U_0$.

\begin{figure}[tb]
    \begin{center}
    \includegraphics[width=3.17in]{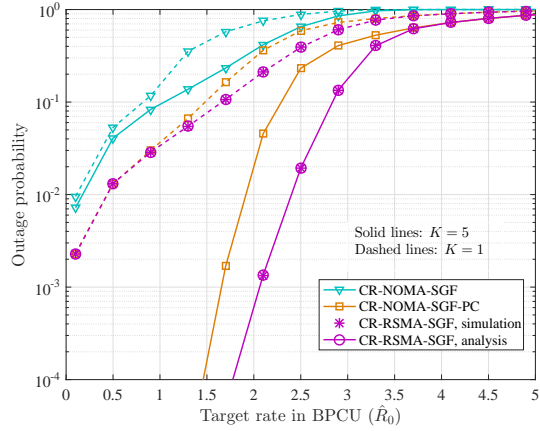}
    \caption{Impact of the target rate on the outage probability.}
    \label{fig:subfig1}
    \end{center}
    \vspace{-0.35in}
\end{figure}

\begin{figure}[tb]
    \begin{center}
    \includegraphics[width=3.17in]{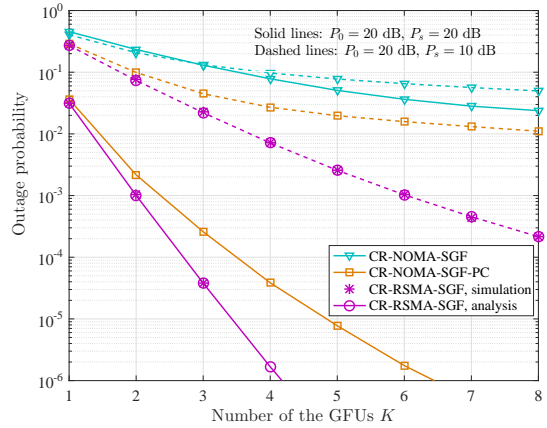}
    \caption{Impact of the number of the GFUs on the outage probability ($\hat R_0 = 1.5$ BPCU and $\hat R_s = 2$ BPCU).}
    \label{fig:subfig3c}
    \end{center}
    \vspace{-0.35in}
\end{figure}

In Fig. 5, we examine the accuracy of the derived analytical expressions for the outage probability. In Fig. 5, we set $\hat R_0 = 2$ BPCU, $\hat R_s = 1.5$ BPCU, and $P_s = \frac{P_0}{15}$. For the ``approximation I'', we use the expression  provided in Theorem 2. For the ``approximation II'', we use the expressions  provided in Corollaries 1 and 2 for $K>2$ and $K=1$, respectively.
The curves in Fig. 5 verify that the accuracy of derived analytical expression in Theorem 1. For the approximated expressions, we can see that the curves of the ``approximation I'' match well with the exact results in the high SNR region, whereas the curves of the ``approximation II'' matches well with the exact results in the high SNR region only for small $K$ values. For large $K$ values ($K = 4$ in this example), a gap exists between the curves of  the ``approximation II'' and exact results. The reason for this phenomenon is that we ignore the terms being proportional to $\frac{1}{P_s^{K+1}}$ and  $\frac{1}{P_s^{K+2}}$ in the expression in Corollary 1.

The impact of the target  rate on the outage probability is investigated in Fig. 6. For the simulation results in Fig. 6, we set $P_0 = 10$ dB, $P_s = 15$ dB, and $\hat R_0 = \hat R_s$. It is seen in Fig. 6 that for given transmit SNR values, the CR-RSMA-SGF transmissions  achieve the smallest outage probabilities in the considered whole target rate region. As the target  rate increases, the outage probability values achieved by all the SGF schemes increase and approach 1.

The impact of the number of the GFUs on the outage probability is investigated in Fig. 7. For the simulation corresponding to Fig. 7, we set \{$\hat R_0 = 1.5$ BPCU, $\hat R_s = 2$ BPCU\} and consider two cases of transmit SNR settings \{$P_0 = 20$ dB, $P_s = 10$ dB\} and \{$P_0 = 20$ dB, $P_s = 10$ dB\}. As $K$ increases, the outage probabilities achieved by the three SGF schemes decreases. Furthermore, Fig. 7 verifies that the multiuser diversity can be achieved by the CR-RSMA-SGF transmissions.
Thus, among the three SGF schemes, the CR-RSMA-SGF transmissions achieve the smallest outage probability for different numbers of the GFUs.

\section{Conclusions}

In this paper, we have proposed a new CR-RSMA to improve the outage performance of SGF transmissions. By applying rate-splitting under the CR principle, the CR-RSMA-SGF transmissions can effectively utilize the transmit power to attain the maximum achievable rate for the admitted GFU. Without introducing intolerable interference to the GBU, the CR-RSMA-SGF transmissions have significantly
extended the non-outage zone compared to the CR-NOMA-SGF transmissions. We have derived exact and approximated expressions for the outage probability of the admitted GFU and revealed that the full multiuser diversity gain can be achieved. Simulation results have verified the superior outage performance achieved by the CR-RSMA-SGF transmissions.

\section*{Appendix A: A proof of Theorem 1}
\renewcommand{\theequation}{A.\arabic{equation}}
\setcounter{equation}{0}

To derive the admitted GFU's outage probability, we evaluate the probability terms $P_{\rm out}^{(\rm I)}$, $P_{\rm out}^{(\rm II)}$, and $P_{\rm out}^{(\rm III)}$, respectively.
In Case II,  $P_{\rm out}^{(\rm II)} =  \sum\nolimits_{k=0}^{K-1} P_{\rm out}^{(\rm II,\it{k})} $ is derived by evaluating the different $P_{\rm out}^{(\rm II,\it{k})}$ taking into account the corresponding order statistics.

\subsection{Evaluation of $P_{\rm out}^{(\rm II,0)}$}

To evaluate $P_{\rm out}^{(\rm II,0)}$, let us introduce $S_0$ as
\begin{eqnarray}
S_0 = \Pr \left( |h_1|^2 > \frac{P_0\epsilon_0^{-1}|h_0|^2 - 1 }{P_s}, |h_K|^2 < \frac{(1+\epsilon_s)(1+\epsilon_0)-(1 + P_0|h_0|^2 ) }{P_s}  \right).  \label{ap:S_0}
\end{eqnarray}
Then, $P_{\rm out}^{(\rm II,0)}$ in \eqref{eq:Pout_II_0} can be rewritten as
\begin{eqnarray}
    P_{\rm out}^{(\rm II,0)} &\!\!\!=\!\!\!&
    \mathop{\mathbb{E}}\limits_{ \eta_0  <   |h_0|^2 < \eta_0(1+\epsilon_s) + \frac{\epsilon_s}{P_0}   }  \left\{ S_0 \right\},   \label{ap:S_0}
\end{eqnarray}
where $\mathop{\mathbb{E}}\{\cdot \}$ stands for the expectation operation, $\epsilon_0 = 2^{\hat R_0}-1$, $\epsilon_s = 2^{\hat R_s} - 1  $, and $\eta_0 = \frac{\epsilon_0}{P_0}$.
In \eqref{ap:S_0}, the expectation on $S_0$ is conducted over $\eta_0 < |h_0|^2 < \eta_0  (1+\epsilon_s) + \frac{\epsilon_s}{P_0}$ taking into account $\hat \tau > 0$ and $ \frac{(1+\epsilon_s)(1+\epsilon_0)-(1 + P_0|h_0|^2 ) }{P_s}$  $> 0 $.
Since the upper bound on $|h_K|^2$ should be larger than the lower bound on $|h_1|^2$, the expectation in \eqref{ap:S_0} should consider the  hidden constraint
$ |h_0|^2 < \eta_0 (1 + \epsilon_s)
$ as well,
so that $P_{\rm out}^{(\rm II,0)}$ is rewritten as
\begin{eqnarray}
    P_{\rm out}^{(\rm II,0)} =
    \mathop{\mathbb{E}}\limits_{ \eta_0  <   |h_0|^2 < \eta_0(1+\epsilon_s)    }  \left\{ S_0 \right\} .  \label{ap:Pout_II_0}
\end{eqnarray}
The joint probability density function (PDF) of the order statistics  $|h_1|^2$ and $|h_K|^2$ is given by \cite{Order-statistics:David}
\begin{eqnarray}
    f_{|h_1|^2, |h_K|^2}(x,y) = \varphi_0 e^{-x} (e^{-x}-e^{-y})^{K-2} e^{-y},
\end{eqnarray}
where $x < y$ and $\varphi_0 = \frac{K!}{(K-2)!}$. Then, $S_0$ can be evaluated as follows:
\begin{eqnarray}
    S_0  &\!\!\!=\!\!\!& \varphi_0  \sum\limits_{i=0}^{K-2}  \binom{K\!-\!2}{i}  (-1)^i \int\nolimits_{\frac{\eta_0^{-1}|h_0|^2 - 1 }{P_s}}^{\frac{(1+\epsilon_s)(1+\epsilon_0)-(1 + P_0|h_0|^2 ) }{P_s}}  e^{-(K-i-1)x} \nonumber \\
    &\!\!\! \!\!\!& \times   \int\nolimits_{x}^{\frac{(1+\epsilon_s)(1+\epsilon_0)-(1 + P_0|h_0|^2 ) }{P_s}}  e^{-(i+1)y}  dydx \nonumber \\
    &\!\!\!=\!\!\!& \varphi_0  \sum\limits_{i=0}^{K-2}  \binom{K\!-\!2}{i}  \frac{(-1)^i}{i + 1}  \nonumber \\
    &\!\!\! \!\!\!& \times \left( \frac{\tilde\mu_3 e^{-\tilde\mu_4 |h_0|^2} -  \tilde\mu_5 e^{-\tilde\mu_{6} |h_0|^2}}{K}   - \frac{\tilde\mu_{1} e^{-\tilde\mu_{2} |h_0|^2} -  \tilde\mu_5 e^{-\tilde\mu_{6} |h_0|^2}}{K - i - 1}    \right), \label{ap:S0}
\end{eqnarray}
where $\tilde \mu_{1} = e^{\frac{K - (1+i)(1+\epsilon_0)(1+\epsilon_s)}{P_s}}$, and $\tilde \mu_{2} = \frac{K-i-1}{P_s \eta_0} - \frac{P_0(1+i)}{P_s}$,
$\tilde \mu_3 = e^{\frac{K}{P_s}}$, $\tilde \mu_4 = \frac{K}{P_s \eta_0}$, $\tilde \mu_5 = e^{-\frac{K(\eta_0 + \eta_s + \eta_0 \eta_s)}{P_s}}$, and $\tilde \mu_{6} = - \frac{KP_0}{P_s}$, and $\eta_s = \frac{\epsilon_s}{P_s}$.

Next, we introduce a  term $\nu(i, \mu)$ as follows:
\begin{eqnarray}
    \nu(i, \mu)  &\!\!\!\triangleq \!\!\!&  \mathop{\mathbb{E}}\limits_{\eta_0 < |h_0|^2 < \eta_0  (1+\epsilon_s)   } \left\{  e^{-\left(\frac{i  }{P_s\eta_0}  + \mu \right) |h_0|^2 } \right\}  \nonumber \\
    &\!\!\!=\!\!\!& \int\nolimits_{\eta_0 }^{\eta_0 (1+\epsilon_s) }  e^{-\left(\frac{i  }{P_s\eta_0}  + \mu + 1\right) x } dx  \nonumber \\
    &\!\!\!=\!\!\!&
   \left\{ {\begin{array}{*{20}{c}}
        {\epsilon_s  \eta_0  }, &{ {\rm if~~} \mu = -1 - \frac{i}{P_s\eta_0} },\\
        {\frac{e^{- \eta_0 \left(\frac{i  }{P_s\eta_0}  + \mu + 1\right) } - e^{- \eta_0 (1+\epsilon_s)  \left(\frac{i  }{P_s\eta_0}  + \mu + 1\right)  }}{  \frac{i }{P_s\eta_0}  + \mu + 1  } }, &{\text{ otherwise}}.
        \end{array}} \right.    \label{ap:nu_term}
\end{eqnarray}
By substituting  \eqref{ap:nu_term} into \eqref{ap:S0}, $P_{\rm out}^{(\rm II,0)}$ can be evaluated as
\begin{eqnarray}
    P_{\rm out}^{(\rm II,0)} &\!\!\!=\!\!\!& \varphi_0  \sum\limits_{i=0}^{K-2}  \binom{K\!-\!2}{i}  \frac{(-1)^i}{i + 1}  \left( \frac{\tilde \mu_3 \nu(0, \tilde \mu_4) -  \tilde \mu_5 \nu(0,\tilde \mu_{6})}{K}   - \frac{\tilde \mu_{1} \nu(0,\tilde \mu_{2})  -  \tilde \mu_5 \nu(0,\tilde \mu_{6}) }{K - i - 1}  \right) \nonumber \\
    &\!\!\!\mathop  = \limits^{(a)}\!\!\!&\! -\frac{\varphi_0}{K\!-\!1} \! \sum\limits_{n=0}^{K-1} \! \binom{K\!-\!1}{n}  \!(-1)^n \! \left( \frac{\tilde \mu_3 \nu(0, \tilde \mu_4) \!-\!  \tilde \mu_5 \nu(0, \tilde \mu_{6})}{K}   \!-\! \frac{\mu_{1} \nu(0,\mu_{2})  \!-\!  \tilde \mu_5 \nu(0,\tilde \mu_{6}) }{K - n}    \right)\!,
     ~~~~~  \label{ap:S_0_2}
\end{eqnarray}
where $\mu_{1} = e^{\frac{K - n(1+\epsilon_0)(1+\epsilon_s)}{P_s}}$ and $\mu_{2} = \frac{K-n}{P_s \eta_0} - \frac{P_0 n}{P_s}$. In step (a) of \eqref{ap:S_0_2}, we have applied $\binom{K-2}{i} = \binom{K-1}{i+1} \frac{i+1}{K-1}$, replaced $n = i+ 1$, and  added the term for $n=0$ without changing the summation since $\frac{\tilde \mu_3 \nu(0, \tilde \mu_4) \!-\!  \tilde \mu_5 \nu(0, \tilde \mu_{6})}{K}   \!-\! \frac{\mu_{1} \nu(0,\mu_{2})  \!-\!  \tilde \mu_5 \nu(0,\tilde \mu_{6}) }{K - n} =0$ when $n = 0$.

By eliminating the terms that are independent of $n$ using $\sum\nolimits_{n=0}^{K-1}  \binom{K-1}{n}  (-1)^n = 0$, \eqref{ap:S_0_2} can be simplified as:
\begin{eqnarray}
    P_{\rm out}^{(\rm II,0)} &\!\!\!=\!\!\!& \frac{\varphi_0}{K-1}  \sum\limits_{n=0}^{K-1}  \binom{K\!-\!1}{n}   (-1)^n  ~ \frac{\mu_{1} \nu(0, \mu_{2})  -  \tilde \mu_5 \nu(0,\tilde\mu_{6}) }{K - n}  \nonumber \\
    &\!\!\!=\!\!\!&   \frac{\varphi_0}{K(K-1)}  \sum\limits_{n=0}^{K}  \binom{K}{n}   (-1)^n  \left( \mu_{1} \nu(0,  \mu_{2})  -  \tilde \mu_5 \nu(0,\tilde \mu_{6}) \right)  , \label{ap:Q_0_4}
\end{eqnarray}
where the term for $n = K$ is added without changing the summation since $ \mu_{1} \nu(0,  \mu_{2})  -  \tilde \mu_5 \nu(0,\tilde \mu_{6}) = 0$ when  $n = K$.
Again, by eliminating the terms that are independent of $n$ using $\sum\nolimits_{n=0}^{K}  \binom{K}{n}  (-1)^n = 0$, $P_{\rm out}^{(\rm II,0)}$ can be further simplified as:
\begin{eqnarray}
    P_{\rm out}^{(\rm II,0)} &\!\!\!=\!\!\!& \frac{\varphi_0}{K(K-1)}  \sum\limits_{n=0}^{K}  \binom{K}{n}   (-1)^n  ~ \mu_{1} \nu(0, \mu_{2})  . \label{ap:Q_0_5}
\end{eqnarray}

 \subsection{Evaluation of $P_{\rm out}^{(\rm II,\it{k})}$ with $1 \le k \le K-2$}

 When $1 \le k \le K-2$, three order statistics, $h_k$, $h_{k+1}$, and $h_K$, are involved in  $P_{\rm out}^{(\rm II,\it{k})}$, which can be rewritten as:
 \begin{eqnarray}
     P_{\rm out}^{(\rm II,\it{k})} &\!\!\!=\!\!\!& \Pr \left( |h_0|^2 > \eta_0,  |h_k|^2 < \frac{\tau}{P_s}, |h_{k+1}|^2 > \frac{\tau}{P_s} , R_{_{{\rm II},K}}  < \hat R_k \right) \nonumber \\
     &\!\!\!=\!\!\!&   \mathop{\mathbb{E}}\limits_{\eta_0 < |h_0|^2 < \eta_0 (1+\epsilon_s)   } \{ S_k \}, \label{ap:S_k}
 \end{eqnarray}
 where $S_k$ is defined by
 \begin{eqnarray}
     S_k &\!\!\! \triangleq \!\!\!& \Pr \left(  |h_k|^2 < \frac{P_0\epsilon_0^{-1}|h_0|^2 - 1 }{P_s}, |h_{k+1}|^2 > \frac{P_0\epsilon_0^{-1}|h_0|^2 - 1 }{P_s} , \right.  \nonumber \\
     &\!\!\! \!\!\!& ~~~~~   \left.  |h_K|^2 < \frac{(1+\epsilon_s)(1+\epsilon_0)-(1 + P_0|h_0|^2 ) }{P_s}  \right). \label{ap:S_k_0}
 \end{eqnarray}
 In \eqref{ap:S_k}, the expectation is  taken over $\eta_0 < |h_0|^2 < \eta_0 (1+\epsilon_s) $ considering the relationship between the upper and lower bounds on the channel gains.

The joint PDF of three order statistics, $h_k$, $h_{k+1}$, and $h_K$, is given by \cite{Order-statistics:David}
\begin{eqnarray}
    f_{|h_k|^2, |h_{k+1}|^2, |h_K|^2} (x, y, z) &\!\!\!\!=\!\!\!\!& \tilde \varphi_k e^{-x} (1-e^{-x})^{k-1} e^{-y} (e^{-y} - e^{-z})^{K-k-2} e^{-z} \nonumber \\
    &\!\!\!\!=\!\!\!\!& \tilde \varphi_k \!\! \sum\limits_{i=0}^{\!K\!-k\!-\!2\!} \!\! \binom{K\!-\!k\!-\!2}{i} \! (-1)^i e^{-x} (1-e^{-x})^{k-1} e^{-(K-k-i-1)y} e^{-(i+1)z}, ~~~~~~~ \label{ap:pdf3}
\end{eqnarray}
where $x \le y \le z$ and $\tilde \varphi_k = \frac{K!}{(k-1)! (K-k-2)!}$. Using \eqref{ap:pdf3}, $S_k$ can be expressed in terms of $|h_0|^2$ as follows:
\begin{eqnarray}
    S_k  &\!\!\!=\!\!\!&\tilde \varphi_k \! \sum\limits_{i=0}^{K-k-2} \!\! \binom{K\!-\!k\!-\!2}{i}  (-1)^i \int\nolimits_{0}^{\frac{\eta_0^{-1}|h_0|^2 - 1 }{P_s}} e^{-x} (1-e^{-x})^{k-1} \nonumber \\
    &\!\!\! \!\!\!& \times \int\nolimits_{\frac{\eta_0^{-1}|h_0|^2 - 1 }{P_s}}^{\frac{(1+\epsilon_s)(1+\epsilon_0)-(1 + P_0|h_0|^2 ) }{P_s}}  e^{-(K-k-i-1)y}  \int\nolimits_{y}^{\frac{(1+\epsilon_s)(1+\epsilon_0)-(1 + P_0|h_0|^2 ) }{P_s}}  e^{-(i+1)z}  dzdydx . ~~~~
\end{eqnarray}
After some algebraic manipulations, $S_k$ can be further evaluated as follows:
\begin{eqnarray}
    S_k  &\!\!\!=\!\!\!& \tilde \varphi_k \! \sum\limits_{i=0}^{K-k-2} \!\! \binom{K\!-\!k\!-\!2}{i} \sum\limits_{\ell=0}^{k} \binom{k}{\ell} \frac{(-1)^{\ell+i}e^{\frac{\ell}{P_s}} e^{-\frac{\ell|h_0|^2  }{P_s\eta_0} }}{k (i+1)} \nonumber \\
    &\!\!\! \!\!\!& \times \left( \frac{\tilde \mu_1 e^{-\tilde \mu_2 |h_0|^2} -  \tilde \mu_5 e^{-\tilde \mu_6 |h_0|^2}}{K - k}   - \frac{\tilde \mu_3 e^{-\tilde \mu_4 |h_0|^2} -  \tilde \mu_5 e^{-\tilde \mu_6 |h_0|^2}}{K - k - i - 1}    \right), \label{ap:S_k_2}
\end{eqnarray}
where $\tilde \mu_1 = e^{\frac{K-k}{P_s}}$, $\tilde \mu_2 = \frac{K-k}{P_s\eta_0}$,  $\tilde \mu_3 = e^{\frac{K-k - (1+i) \left(1+\epsilon_0\right) \left(1+ \epsilon_s\right) }{P_s}}$, $\tilde \mu_4 = \frac{K-k-i-1}{P_s \eta_0 }  - \frac{(1+i)P_0}{P_s}  $,
$\tilde \mu_5 = e^{-\frac{(K-k)\left(\epsilon_0 + \epsilon_s + \epsilon_0\epsilon_s  \right)}{P_s }}$, and $\tilde \mu_6 = -\frac{(K - k)P_0 }{P_s } $.

By substituting \eqref{ap:S_k_2} into
\eqref{ap:S_k}, $P_{\rm out}^{(\rm II,\it{k})}$ can be evaluated as:
\begin{eqnarray}
    P_{\rm out}^{(\rm II,\it{k})}  &\!\!\!=\!\!\!& \tilde \varphi_k \! \sum\limits_{i=0}^{K-k-2} \!\! \binom{K\!-\!k\!-\!2}{i} \frac{(-1)^i }{k (i+1)} \sum\limits_{n=0}^{k} \binom{k}{n} (-1)^{n}e^{\frac{n}{P_s}} \nonumber \\
    &\!\!\! \!\!\!& \times \left( \frac{\tilde \mu_1 \nu(n, \tilde \mu_2)  -  \tilde \mu_5 \nu(n, \tilde \mu_6)}{K - k}   - \frac{\tilde \mu_3 \nu(n, \tilde \mu_4) -  \tilde \mu_5 \nu(n, \tilde \mu_6) }{K - k - i - 1}    \right)   \nonumber \\
    &\!\!\! \mathop  = \limits^{(a)} \!\!\!&
    \frac{-\tilde \varphi_k}{ k (K-k-1)} \! \sum\limits_{m=0}^{K\!-\!k\!-\!1} \!\! \binom{K-k-1}{m} (-1)^m \sum\limits_{n=0}^{k} \binom{k}{n} (-1)^{n}e^{\frac{n}{P_s}} \nonumber \\
    &\!\!\! \!\!\!& \times \left( \frac{\tilde \mu_1 \nu(n, \tilde \mu_2)  -  \tilde \mu_5 \nu(n, \tilde \mu_6)}{K - k}   - \frac{ \mu_3 \nu(n,  \mu_4) -  \tilde \mu_5 \nu(n, \tilde \mu_6) }{K - k - m}    \right),
    \label{ap:S_k_3}
\end{eqnarray}
where $ \mu_3 = e^{\frac{K-k - m \left(1+\epsilon_0 \right) \left(1+ \epsilon_s \right) }{P_s}}$  and $ \mu_4 = \frac{K-k-m}{P_s \eta_0 }  - \frac{m P_0}{P_s}  $. In step (a) of \eqref{ap:S_k_3}, we have replaced $\binom{K-k-2}{i}$ with $\binom{K-k-1}{i+1} \frac{i+1}{K-k-1}$, applied $m = i+ 1$, and added term for $m=0$ without changing the value of $P_{\rm out}^{(\rm II,\it{k})}$ since $\frac{\tilde \mu_1 \nu(n, \tilde \mu_2)  -  \tilde \mu_5 \nu(n, \tilde \mu_6)}{K - k}   - \frac{ \mu_3 \nu(n,  \mu_4) -  \tilde \mu_5 \nu(n, \tilde \mu_6) }{K - k - m} = 0$ when $m=0$.

Furthermore, some terms in \eqref{ap:S_k_3} involving $\tilde \mu_1$, $\tilde \mu_2$, $\tilde \mu_5$, and $\tilde \mu_6$ but being independent of $m$ can be further eliminated since $\sum\nolimits_{m=0}^k \binom{k}{m} (-1)^m = 0$, while $\tilde \mu_1$, $\tilde \mu_2$, $\tilde \mu_5$, and $\tilde \mu_6$ are not functions of $m$. The simplification can be expressed as follows:
\begin{eqnarray}
    P_{\rm out}^{(\rm II,\it{k})}  &\!\!\!=\!\!\!& \frac{\tilde \varphi_k}{ k (K-k-1)} \! \sum\limits_{m=0}^{K-k-1} \!\! \binom{K\!-\!k\!-\!1}{m} (-1)^m \sum\limits_{n=0}^{k} \binom{k}{n} (-1)^{n}e^{\frac{n}{P_s}}   \frac{ \mu_3 \nu(n,  \mu_4) -  \tilde \mu_5 \nu(n, \tilde \mu_6) }{K - k - m}   \nonumber \\
    &\!\!\! \mathop  = \limits^{(a)} \!\!\!&   \varphi_k  \sum\limits_{m=0}^{K-k}   \binom{K\!-\!k}{m} (-1)^m \sum\limits_{n=0}^{k} \binom{k}{n} (-1)^{n}e^{\frac{n}{P_s}}  ( \mu_3 \nu(n, \mu_4) -  \tilde \mu_5 \nu(n, \tilde \mu_6))  \nonumber \\
    &\!\!\! \mathop  = \limits^{(b)} \!\!\!&  \varphi_k  \sum\limits_{m=0}^{K-k}   \binom{K\!-\!k}{m} (-1)^m \sum\limits_{n=0}^{k} \binom{k}{n} (-1)^{n}e^{\frac{n}{P_s}}  \mu_3 \nu(n,   \mu_4) ,
\end{eqnarray}
where step (a) follows by absorbing $K-k-1$ into the binomial coefficients without changing the summation and step (b) follows by using $\sum\nolimits_{m=0}^k \binom{k}{m} (-1)^m = 0$  and $\tilde \mu_5$  and $\tilde \mu_6$ that are not functions of $m$.

\subsection{Evaluation of $P_{\rm out}^{(\rm II,{\it{K}}-1)}$}

The probability term $P_{\rm out}^{(\rm II,{\it{K}}-1)}$ can be expressed as:
\begin{eqnarray}
    P_{\rm out}^{(\rm II,{\it{K}}-1)} &\!\!\!=\!\!\!&  \mathop{\mathbb{E}}\limits_{|h_0|^2 > \eta_0 }  \left\{  \Pr \left( |h_{K-1}|^2 < \frac{P_0\epsilon_0^{-1}|h_0|^2 - 1 }{P_s}, |h_K|^2 > \frac{P_0\epsilon_0^{-1}|h_0|^2 - 1 }{P_s} , \right. \right.   \nonumber \\
    &\!\!\! \!\!\!&    \left. \left.  |h_K|^2 < \frac{(1+\epsilon_s)(1+\epsilon_0)-(1 + P_0|h_0|^2 ) }{P_s}  \right) \right\} .  \label{ap:S_K11}
\end{eqnarray}
By extracting the hidden constraint on the upper and lower bounds on $|h_K|^2$ from \eqref{ap:S_K11}, i.e., $ \frac{P_0\epsilon_0^{-1}|h_0|^2 - 1 }{P_s} < \frac{(1+\epsilon_s)(1+\epsilon_0)-(1 + P_0|h_0|^2 ) }{P_s}$, $P_{\rm out}^{(\rm II,{\it{K}}-1)}$ can be rewritten as follows:
\begin{eqnarray}
    P_{\rm out}^{(\rm II,{\it{K}}-1)} &\!\!\!=\!\!\!&  \mathop{\mathbb{E}}\limits_{\eta_0 < |h_0|^2 < (1+\epsilon_s) \eta_0  } \{ S_{K-1} \},  \label{ap:Q_k_11}
\end{eqnarray}
where $S_{K-1}$ denotes probability inside the expectation in \eqref{ap:S_K11}. There are two order statistics $h_{K-1}$ and $h_{K}$ involving in $S_{K-1}$ with the  joint PDF \cite{Order-statistics:David}
\begin{eqnarray}
    f_{|h_{K-1}|^2, |h_K|^2} (x, y) = \varphi_0 e^{-x} (1-e^{-x})^{K-2} e^y,  \label{ap:PDF_3}
\end{eqnarray}
where $x \le y$. Using \eqref{ap:PDF_3}, $S_{K-1}$ can be evaluated as follows:
\begin{eqnarray}
    S_{K-1} = \varphi_0 \sum\limits_{n=0}^{K-1} \binom{K\!-\!1}{n} \frac{(-1)^n e^{\frac{n}{P_s} } e^{-\frac{n|h_0|^2  }{P_s\eta_0} } }{K-1}  \left( e^{\frac{1}{P_s}} e^{-\mu_5 |h_0|^2}  - e^{-\frac{\epsilon_0 + \epsilon_s + \epsilon_0\epsilon_s  }{P_s} } e^{-\mu_6 |h_0|^2}    \right), ~~
\end{eqnarray}
where $\mu_5 = \frac{1}{P_s\eta_0}$ and $\mu_6 = -\frac{P_0}{P_s}$.
Using the expression in \eqref{ap:nu_term}, $P_{\rm out}^{(\rm II,{\it{K}}-1)}$ can be derived as follows:
\begin{eqnarray}
    P_{\rm out}^{(\rm II,{\it{K}}-1)} = \frac{\varphi_0}{K-1} \sum\limits_{n=0}^{K-1} \binom{K\!-\!1}{n}  (-1)^n e^{\frac{n}{P_s} }    \left( e^{\frac{1}{P_s}} \nu(n, \mu_5)  - e^{-\frac{\epsilon_0 + \epsilon_s + \epsilon_0\epsilon_s  }{P_s} } \nu(n, \mu_6)   \right).
\end{eqnarray}

\subsection{Evaluation of $P_{\rm out}^{(\rm I)}$ and $P_{\rm out}^{(\rm III)}$}

In Case I, the determination of $P_{\rm out}^{(\rm I)}$ involves two independent random variables $|h_0|^2$ and $|h_K|^2$.
Recalling the expression in \eqref{eq:Pout_I}, $P_{\rm out}^{(\rm I)}$ can be rewritten as follows:
\begin{eqnarray}
    P_{\rm out}^{(\rm I)} =  \mathop{\mathbb{E}}\limits_{|h_0|^2 > \eta_0 }  \left\{   \Pr \left( |h_K|^2 < \frac{\eta_0^{-1}|h_0|^2 - 1 }{P_s},   |h_K|^2 < \frac{\epsilon_s }{ P_s }  \right)  \right\} .
\end{eqnarray}
By comparing $\frac{\eta_0^{-1}|h_0|^2 - 1 }{P_s}$ and $ \frac{\epsilon_s }{ P_s }$, it can be seen that $\frac{\eta_0^{-1}|h_0|^2 - 1 }{P_s} < \frac{\epsilon_s }{ P_s }$ if $|h_0|^2  < \eta_0 (1+\epsilon_s)$; Otherwise,  $\frac{\eta_0^{-1}|h_0|^2 - 1 }{P_s} > \frac{\epsilon_s }{ P_s }$.  Thus, $P_{\rm out}^{(\rm I)}$ can be evaluated as follows:
\begin{eqnarray}
    P_{\rm out}^{(\rm I)} &\!\!\!=\!\!\!&  \mathop{\mathbb{E}}\limits_{ \eta_0 < |h_0|^2  < \eta_0 (1+\epsilon_s) }  \left\{   \Pr \left( |h_K|^2 < \frac{\eta_0^{-1}|h_0|^2 - 1 }{P_s} \right)  \right\}  + \mathop{\mathbb{E}}\limits_{  |h_0|^2  > \eta_0 (1+\epsilon_s) }  \left\{   \Pr \left(  |h_K|^2 <  \eta_s   \right)  \right\}  \nonumber \\
    &\!\!\!= \!\!\!& \int\nolimits_{\eta_0}^{ \eta_0 (1+\epsilon_s)} \left(1 - e^{-\frac{\eta_0^{-1}x - 1 }{P_s} } \right)^K e^{-x} dx + \left(1 - e^{-\eta_s}  \right)^K e^{- \eta_0 (1+\epsilon_s)  }      \nonumber \\
    &\!\!\!= \!\!\!& \sum\limits_{n = 0}^K \binom{K}{n} (-1)^n e^{\frac{n}{P_s}} \int\nolimits_{\eta_0 }^{\eta_0 (1+\epsilon_s) }  e^{-\left(\frac{n  }{P_s\eta_0}   + 1\right) x } dx    + \left(1 - e^{-\eta_s}  \right)^K e^{- \eta_0 (1+\epsilon_s)  }    \nonumber \\
    &\!\!\!= \!\!\!&  \sum\limits_{n = 0}^K \binom{K}{n} (-1)^n e^{\frac{n}{P_s}} \nu(n, 0)   + \left(1 - e^{-\eta_s}  \right)^K e^{- \eta_0 (1+\epsilon_s)  }.
\end{eqnarray}

Similarly to $P_{\rm out}^{(\rm I)}$, $P_{\rm out}^{(\rm III)}$ is a function of two independent random variables $|h_0|^2$ and $|h_K|^2$. In Case III, the achievable rate is $R_{K}^{(\rm III)} =  \log_2\left( 1 + \frac{P_s|h_K|^2 }{P_0|h_0|^2  + 1} \right)$ and $\tau = 0$. Then, $P_{\rm out}^{(\rm III)}$ can be evaluated as follows:
\begin{eqnarray}
    P_{\rm out}^{(\rm III)} &\!\!\!= \!\!\!&  \Pr \left( |h_0|^2 < \eta_0,  \log_2\left( 1 + \frac{P_s|h_K|^2 }{P_0|h_0|^2  + 1} \right) < \hat R_s  \right)  \nonumber \\
    &\!\!\!= \!\!\!& \sum\limits_{n = 0}^K \binom{K}{n}  (-1)^n e^{-n \eta_s} \frac{1 - e^{-(1+n \eta_s P_0 )\eta_0 }}{1 + n \eta_s  P_0}.
\end{eqnarray}

By combing the obtained expressions for $P_{\rm out}^{(\rm I)}$, $P_{\rm out}^{(\rm II)}$, and $P_{\rm out}^{(\rm III)}$, we arrive at \eqref{eq:Pout_GFU}.

\section*{Appendix B: A proof of Theorem 2}
\renewcommand{\theequation}{B.\arabic{equation}}
\setcounter{equation}{0}
\setcounter{subsection}{0}

Considering that $P_{\rm out}^{(\rm I)}$, $P_{\rm out}^{(\rm II)}$, and $P_{\rm out}^{(\rm III)}$ depend on $k$, the high SNR approximations for $P_{\rm out}^{(\rm I)}$, $P_{\rm out}^{(\rm II)}$, and $P_{\rm out}^{(\rm III)}$ will be derived separately in the following subsections.

\subsection{High SNR approximation for $P_{\rm out}^{(\rm II,0)}$}

Based on the derived closed-form expression in \eqref{ap:Q_0_5}, $P_{\rm out}^{(\rm II,0)}$ can be rewritten as follows:
\begin{eqnarray}
    P_{\rm out}^{(\rm II,0)} &\!\!\!=\!\!\!& \frac{\varphi_0}{K(K-1)}  \sum\limits_{n=0}^{K}  \binom{K}{n}   (-1)^n   \mu_{1} \nu(0, \mu_{2}) . \nonumber \\
    &\!\!\!=\!\!\!& \frac{\varphi_0}{K(K-1)}  \sum\limits_{n=0}^{K}  \binom{K}{n}   (-1)^n \mu_1 \int\nolimits_{\eta_0}^{\eta_0(1+\epsilon_s)} e^{-(\mu_2+1)x} dx  . \label{apb:Q_0}
\end{eqnarray}
By applying the approximation $e^{-x} = 1 - x$ for $x \to 0$ and using the definitions of $\mu_1$ and $\mu_2$, as $P_0 = P_s \to \infty$, $P_{\rm out}^{(\rm II,0)}$ can be approximated as follows:
\begin{eqnarray}
    P_{\rm out}^{(\rm II,0)} &\!\!\!=\!\!\!& \frac{\varphi_0}{K(K-1)}  \int\nolimits_{\eta_0}^{\eta_0(1+\epsilon_s)} \sum\limits_{n=0}^{K}  \binom{K}{n}   (-1)^n  e^{-\frac{n (1+\epsilon_0 )(1+ \epsilon_s)}{P_s}}  e^{n (\epsilon_0^{-1}  + 1  )x}  dx   \nonumber \\
    &\!\!\! \mathop = \limits^{(a)} \!\!\!&  \frac{\varphi_0}{K(K-1)}  \int\nolimits_{\eta_0}^{\eta_0(1+\epsilon_s)}  \left( 1  -  e^{-\left( \frac{ (1+\epsilon_0 )(1+ \epsilon_s)}{P_s} - (\epsilon_0^{-1}  +1) x  \right) } \right)^K  dx  \nonumber \\
    &\!\!\!=\!\!\!&    \frac{\varphi_0}{K(K-1)}  \int\nolimits_{\eta_0}^{\eta_0(1+\epsilon_s)} \left(  \frac{ (1+\epsilon_0 )(1+ \epsilon_s)}{P_s} - \left(\epsilon_0^{-1}  +1 \right) x   \right)^K dx,
\end{eqnarray}
where step (a) is obtained by applying $\sum\nolimits_{n}^K \binom{K}{n} (-1)^n a^n  =  (1-a)^K$. By applying the binomial expansion, $P_{\rm out}^{(\rm II,0)}$ can be further simplified as follows:
\begin{eqnarray}
    P_{\rm out}^{(\rm II,0)} &\!\!\!=\!\!\!&  \frac{\varphi_0}{K(K-1)}  \sum\limits_{n = 0}^K \binom{K}{n} \left( \frac{(1+\epsilon_0)(1+\epsilon_s)}{P_s} \right)^{K-n} (-1)^n \left(\epsilon_0^{-1}  + 1  \right)^n   \int\nolimits_{\eta_0}^{\eta_0(1+\epsilon_s)}    x^n dx  \nonumber \\
    &\!\!\!=\!\!\!&   \frac{\varphi_0 \epsilon_0 (1+\epsilon_0)^K}{P_s^{K+1}K(K-1)}   \sum\limits_{n = 0}^K \binom{K}{n} \frac{(-1)^n}{n+1}    \left( (1+\epsilon_s)^{K + 1}  -  (1+\epsilon_s)^{K-n}  \right).   \label{app:Q0}
\end{eqnarray}

\subsection{High SNR approximation for $P_{\rm out}^{(\rm II,\it{k})}$ with $1 \le k \le K-2$}

Reflecting to the derivations made in Appendix A, $P_{\rm out}^{(\rm II,\it{k})}$ can be rewritten as follows:
\begin{eqnarray}
    P_{\rm out}^{(\rm II,\it{k})}  &\!\!\!=\!\!\!& \varphi_k  \sum\limits_{m=0}^{K-k}   \binom{K\!-\!k}{m} (-1)^m \sum\limits_{n=0}^{k} \binom{k}{n} (-1)^{n}e^{\frac{n}{P_s}}  \mu_3
    \int\nolimits_{\eta_0}^{\eta_0(1+\epsilon{_k})}  e^{-\left( \frac{n}{P{_s} \eta_0} + \mu_4 +1 \right)x}  dx.
\end{eqnarray}
By applying the approximation $e^{-x} \approx 1 - x$ for $x \to 0$ and using the definitions of $\mu_3$ and $\mu_4$,  as $P_0 = P_s \to \infty$, $P_{\rm out}^{(\rm II,\it{k})}$ can be approximated as follows:
\begin{eqnarray}
    P_{\rm out}^{(\rm II,\it{k})}  &\!\!\!=\!\!\!& \varphi_k  \sum\limits_{m=0}^{K-k}   \binom{K\!-\!k}{m} (-1)^m \sum\limits_{n=0}^{k} \binom{k}{n} (-1)^{n}e^{\frac{n}{P_s}}   e^{-\frac{ m \left(1+\epsilon_0 \right) \left(1+ \epsilon_s \right) }{P_s}}
    \nonumber \\
    &\!\!\! \!\!\!& \times \int\nolimits_{\eta_0}^{\eta_0(1+\epsilon{_k})}  e^{-\left( \frac{n}{P{_k} \eta_0} -  \frac{m}{P_s \eta_0 }  - m \right)x}  dx.
\end{eqnarray}
Since $\sum\limits_{n=0}^{k} \binom{k}{n} (-1)^{n} a^n = (1-a)^k$, $P_{\rm out}^{(\rm II,\it{k})}$ can be further expressed as follows:
\begin{eqnarray}
    P_{\rm out}^{(\rm II,\it{k})} &\!\!\!=\!\!\!& \varphi_k \int\nolimits_{\eta_0}^{\eta_0(1+\epsilon_s)}   \sum\limits_{m=0}^{K-k}   \binom{K\!-\!k}{m} (-1)^m \sum\limits_{n=0}^{k} \binom{k}{n} (-1)^{n}
    \nonumber \\
    &\!\!\! \!\!\!& \times  e^{-n\left(\frac{    (1+\epsilon_0  )  (1+ \epsilon_s  ) }{P_s} -(\epsilon_0^{-1} + 1)x \right)} e^{-n\left( \epsilon_0^{-1}x -  P_s^{-1}  \right)}  dx \nonumber \\
    &\!\!\! =  \!\!\!&  \varphi_k \int\nolimits_{\eta_0}^{\eta_0(1+\epsilon{_k})} \left( 1 -  e^{- \left(\frac{    (1+\epsilon_0  )  (1+ \epsilon_s  ) }{P_s} -(\epsilon_0^{-1} + 1)x \right)}  \right)^{K-k}   \left( 1 -  e^{- \left( \epsilon_0^{-1}x -P_s^{-1}   \right)} \right)^{k}   dx  \nonumber \\
    &\!\!\! \mathop =\limits^{(a)}  \!\!\!&  \varphi_k \int\nolimits_{\eta_0}^{\eta_0(1+\epsilon{_k})}    \left(\frac{    (1+\epsilon_0  )  (1+ \epsilon_s  ) }{P_s} -(\epsilon_0^{-1} + 1)x   \right)^{K-k}  \left(  \epsilon_0^{-1}x - P_s^{-1}   \right) ^{k}   dx ,  \label{apb:Q_k}
\end{eqnarray}
where step (a) follows high SNR approximations. Applying the binomial expansions to \eqref{apb:Q_k}, the high SNR approximation for $P_{\rm out}^{(\rm II,\it{k})}$ can be obtained as follows:
\begin{eqnarray}
    P_{\rm out}^{(\rm II,\it{k})} &\!\!\! = \!\!\!&  \frac{\varphi_k \epsilon_0  (1+\epsilon_0)^{K-k} (-1)^k }{ P_s^{K+1} } \sum\limits_{m=0}^{K-k} \binom{K-k}{m} (-1)^m (1 + \epsilon_s )^{K-k-m} \nonumber \\
    &\!\!\! \!\!\!&    \times  \sum\limits_{n=0}^{k} \binom{k}{n} (-1)^n
    \frac{(1 + \epsilon_s )^{m+n+1} - 1}{m+n+1}.  \label{app:Qk}
\end{eqnarray}

\subsection{High SNR approximation for $P_{\rm out}^{(\rm II,{\it{K}}-1)}$}

First, we rewrite $P_{\rm out}^{(\rm II,{\it{K}}-1)}$ as follows:
\begin{eqnarray}
    P_{\rm out}^{(\rm II,{\it{K}}-1)}  &\!\!\!=\!\!\!& \frac{\varphi_0}{K-1} \sum\limits_{n=0}^{K-1} \binom{K\!-\!1}{n}  (-1)^n e^{\frac{n}{P_s} }    \left( e^{\frac{1}{P_s}} \nu(n, \mu_5)  - e^{-\frac{\epsilon_0 + \epsilon_s + \epsilon_0\epsilon_s  }{P_s} } \nu(n, \mu_6)   \right) \nonumber \\
    &\!\!\! =  \!\!\!& \frac{\varphi_0}{K-1} \sum\limits_{n=0}^{K-1} \binom{K\!-\!1}{n}  (-1)^n e^{\frac{n}{P_s} }  \int\nolimits_{\eta_0}^{\eta_0(1+\epsilon{_k})}  \left(   e^{\frac{1}{P_s}}  e^{-\left( \frac{n}{P{_k} \eta_0} + \mu_5 +1 \right)x}  \right.  \nonumber \\
    &\!\!\!    \!\!\!& \left. - e^{-\frac{\epsilon_0 + \epsilon_s + \epsilon_0\epsilon_s  }{P_s} }  e^{-\left( \frac{n}{P{_k} \eta_0} + \mu_6 +1 \right)x} \right)  dx.
\end{eqnarray}
Since $\sum\limits_{n=0}^{k} \binom{k}{n} (-1)^{n}   = 0$, $P_{\rm out}^{(\rm II,{\it{K}}-1)}$ can be expressed as follows:
\begin{eqnarray}
    P_{\rm out}^{(\rm II,{\it{K}}-1)}  &\!\!\!=\!\!\!& \frac{\varphi_0}{K-1}   \int\nolimits_{\eta_0}^{\eta_0(1+\epsilon_s)} \left( e^{\frac{1}{P_s} - (\epsilon_0^{-1} + 1)x}  -  e^{-\frac{\epsilon_0 + \epsilon_s + \epsilon_0\epsilon_s  }{P_s} }  \right)  \left( 1 - e^{ - n \left(\epsilon_0^{-1} x - P_s^{-1}  \right)}     \right)^{K-1} dx. ~~~~
\end{eqnarray}
By applying the approximations $e^{-x} = 1- x$ for $x \to 0$, as $P_0 = P_s \to \infty$,  $P_{\rm out}^{(\rm II,{\it{K}}-1)}$ can be approximated as follows:
\begin{eqnarray}
    P_{\rm out}^{(\rm II,{\it{K}}-1)}  &\!\!\!=\!\!\!& \frac{\varphi_0}{K-1}   \int\nolimits_{\eta_0}^{\eta_0(1+\epsilon_s)}     \left( \frac{(1+\epsilon_0)(1+\epsilon_s)}{P_s}  - (\epsilon_0^{-1} + 1)x   \right)  \left(  \epsilon_0^{-1} x - P_s^{-1}   \right)^{K-1}    dx \nonumber \\
    &\!\!\!=\!\!\!& \frac{\varphi_0}{K-1}   \int\nolimits_{\eta_0}^{\eta_0(1+\epsilon_s)}   \epsilon_0^{-(K-1)}   \left( \frac{(1+\epsilon_0)(1+\epsilon_s)}{P_s}  - (\epsilon_0^{-1} + 1)x   \right)    \left(x - \frac{\epsilon_0}{P_s}   \right)^{K-1} dx   \nonumber \\
    &\!\!\!=\!\!\!&  \frac{\varphi_0}{K-1}   \int\nolimits_{\eta_0}^{\eta_0(1+\epsilon_s)}   \epsilon_0^{-(K-1)}     \frac{(1+\epsilon_0)(1+\epsilon_s)}{P_s}   \left(x - \frac{\epsilon_0}{P_s}   \right)^{K-1} dx   \nonumber \\
    &\!\!\! \!\!\!&  - \frac{\varphi_0}{K-1}   \int\nolimits_{\eta_0}^{\eta_0(1+\epsilon_s)}   \epsilon_0^{-(K-1)}    (\epsilon_0^{-1} + 1)x      \left(x - \frac{\epsilon_0}{P_s}   \right)^{K-1} dx   \nonumber \\
    &\!\!\!=\!\!\!&   \frac{\varphi_0 \epsilon_0 \epsilon_s^K (1 + \epsilon_0)(1+ \epsilon_s)  }{ P_s^{K+1}K(K-1)} -
    \frac{ \varphi_0  \epsilon_s^K (\epsilon_0^{-1} + 1) (K(1+\epsilon_s) + 1) }{P_s^{K+1}K(K-1)(K+1)} .
    \label{app:QKm1}
\end{eqnarray}

Following similar algebraic manipulations as for deriving the high SNR approximation of $P_{\rm out}^{({\rm II},k)}$ ($1 \le k \le K-1$), $P_{\rm out}^{(\rm I)}$ and $P_{\rm out}^{(\rm III)}$ can be approximated as
\begin{eqnarray}
    P_{\rm out}^{(\rm I)} = \frac{\epsilon_0 \epsilon_s^{K+1} }{P_s^{K+1}(K+1) }  + \frac{\epsilon_s^K}{P_s^K}  - \frac{\epsilon_0 \epsilon_s^K(1+\epsilon_s)}{P_s^{K+1}}  \label{app:QK}
\end{eqnarray}
and
\begin{eqnarray}
    P_{\rm out}^{(\rm III)} =  \frac{\epsilon_s^K \left( (1 + \epsilon_0)^{K+1} -1 \right) }{P_s^{K+1} (K+1)}  - \frac{\epsilon_s^K \left( (\epsilon_0(K+1) -1 )(1+\epsilon_0)^{K+1} +1  \right) }{P_s^{K+2} (K+2) (K+1)}   ,   \label{app:QKp1}
\end{eqnarray}
respectively.

By combing \eqref{app:Q0}, \eqref{app:Qk}, \eqref{app:QKm1}, \eqref{app:QK}, and \eqref{app:QKp1}, the high SNR approximation for $P_{\rm out}$ is obtained as \eqref{eq:Pout_app}.

\begin{balance}
\bibliography{IEEEabrv,IEEE_bib_new}
\end{balance}

\end{document}